\documentclass{ieeeaccess}
\pdfoutput=1
\usepackage{graphicx}
\usepackage{booktabs}
\usepackage{amsmath}
\usepackage{array}
\usepackage{url}
\vol{14}\year{2026}

\begin{document}

\history{Received 5 September 2026, accepted 20 September 2026. Authors' version of the article published in IEEE Access, vol.~14, pp.~148208--148229, 2026. Copyright 2026 The Authors. This work is licensed under a Creative Commons Attribution 4.0 License (CC BY 4.0).}
\doi{10.1109/ACCESS.2026.3736943}

\title{Distributional Validity of a Korean Synthetic Persona Panel: Evidence
From the Korea Media Panel Survey}

\author{\uppercase{Howard Kim}\authorrefmark{1,2} and \uppercase{Keuntae Cho}\authorrefmark{3}}

\address[1]{College of AI Convergence, Seoul Cyber University, Seoul 01133,
South Korea (e-mail: howardkim@iscu.ac.kr)}
\address[2]{Department of Management of Technology, Graduate School,
Sungkyunkwan University, Suwon 16419, South Korea (e-mail: howardkim@skku.edu)}
\address[3]{Department of Systems Management Engineering, Sungkyunkwan
University, Suwon 16419, South Korea}

\markboth
{Kim \headeretal: Distributional Validity of a Korean Synthetic Persona Panel: Evidence From the Korea Media Panel Survey}
{Kim \headeretal: Distributional Validity of a Korean Synthetic Persona Panel: Evidence From the Korea Media Panel Survey}

\corresp{Corresponding author: Keuntae Cho (e-mail: ktcho@skku.edu).}

\begin{abstract}
Large language model (LLM) personas are proposed as survey
respondents, yet validation outside English-speaking contexts is
scarce. We evaluate how well a Korean synthetic persona panel
used to condition Gemini~3.5~Flash and EXAONE reproduces digital and artificial intelligence (AI) service-use distributions
of the Korea Media Panel Survey. About 8{,}000 personas per model
answered eight service-use items and eight attitudinal
constructs; responses were compared with weighted survey estimates. The overall
mean absolute error (MAE) was 14--19~percentage points (pp), with binary
item-mean correlations of 0.70--0.91 across waves. Segment error across five axes was 14--18~pp, with between-group signed-error ranges of 49.6/34.7~pp (Gemini/EXAONE; 39.5/31.2 without the non-comparable teen cells).
Errors were model-specific: an age stereotype (Gemini)
versus an acquiescence-consistent level bias (EXAONE). Generative-AI overestimation was consistent with temporal misalignment; short-form
underestimation was framing-sensitive and persisted under randomized order (both shown for Gemini).
Post-hoc holdout calibration on 30\% of the real data, with the correction form
selected inside the calibration set, cut cell MAE from 18.3/15.1 to
4.9/4.4~pp, yet direct estimation from that subsample was more accurate than the calibrated panel
(3.6~pp), a synthetic-informed shrinkage estimator beat its real-only counterpart by at most 0.7~pp, and the correction did not transfer competitively across waves. The calibrated panel kept an advantage only below roughly 250--860 real responses (at most 2.3~pp over a real-only shrinkage estimator) or, for one model, on unobserved segments. In this setting, synthetic panels are not survey substitutes; their value is
diagnostic.
\end{abstract}

\begin{IEEEkeywords}
Bias correction, distributional validity, Korea Media Panel Survey, large
language models, personas, survey simulation, synthetic data.
\end{IEEEkeywords}

\titlepgskip=-15pt

\maketitle

\section{Introduction}
\IEEEPARstart{P}{opulation} surveys face rising costs and falling response
rates, while policy demand for timely measurement of fast-moving digital
behaviors continues to grow. For technology management, the problem is sharpest at the moment of
adoption: the uptake of an emerging technology must be gauged early in its
diffusion \cite{rogers}, before any survey wave exists from which to
extrapolate. Respondents generated by large language models
(LLMs) promise substantial savings in
survey cost and time: if a language model conditioned on a realistic persona can
answer a questionnaire as a matched human respondent would, panels could be augmented
or partially replaced. Whether synthetic responses reproduce real population distributions has
rarely been validated outside English-speaking settings. Recent Korean studies have compared LLM-generated responses with an online panel designed to represent the South Korean population and with the Seoul Survey, a municipal probability sample~\cite{leemc,parkseoul}. We extend this work by benchmarking two contrasting models against a nationally representative longitudinal media panel, with explicit real-only baselines, subgroup diagnostics, and temporal and calibration tests. This study addresses three
research questions (RQs).
\begin{itemize}
\item \textbf{RQ1 (overall agreement):} How closely does the synthetic panel's
response distribution match the real distribution?
\item \textbf{RQ2 (segment error):} How does error decompose across demographic
segments, and is it concentrated in particular groups?
\item \textbf{RQ3 (calibration):} Does correcting the synthetic panel with a
small amount of real data improve accuracy on held-out real data, and does the
correction transfer across waves? (RQ3 was formulated after the registration and is post-hoc; Sec.~III-H.)
\end{itemize}
Whereas \emph{silicon sampling} (conditioning an LLM on demographic profiles so
that it simulates survey respondents) matched U.S.\ opinion distributions
well~\cite{argyle}, we show that it fails for Korean digital and artificial intelligence (AI)
service use, explain why, and delimit how far it can be corrected. This study makes three contributions.
The first is a quantitative assessment of the validity of synthetic personas
against a nationally representative Korean media panel, benchmarked against
both naive and real-only baselines. That assessment shows that the raw panel
adds no segment-level value. The second is a diagnosis of the structural error
patterns (temporal misalignment, response style, age-stereotype-consistent
slopes, item framing) together with model-specific bias signatures. The third
is a baseline-disciplined assessment of signature-matched small-sample
calibration. With the correction
form selected inside the calibration set, the correction cuts the held-out sex-by-age cell error by
about 70\%. A real-only estimator is nevertheless more accurate once a few
hundred real responses are available (direct estimation beyond roughly 430--1{,}730 responses; a real-only shrinkage estimator beyond roughly 250--860), a synthetic-informed shrinkage estimator improves on its real-only counterpart by at most 0.7~percentage points (pp), and the correction, although it reduces raw error across waves, does not beat simple target-year and prior-wave references. The calibrated
panel's practical niche is therefore confined to small calibration samples
and (for one model) entirely unobserved segments; below a few hundred responses the margin over a real-only shrinkage estimator is at most about 2~pp.
The remainder of the paper reviews related work (Sec.~II), describes the data,
generation, and validation design (Sec.~III), reports the results (Sec.~IV),
and discusses implications and limitations (Secs.~V and VI).

\Figure[t!](topskip=0pt, botskip=0pt, midskip=0pt)[width=\textwidth]{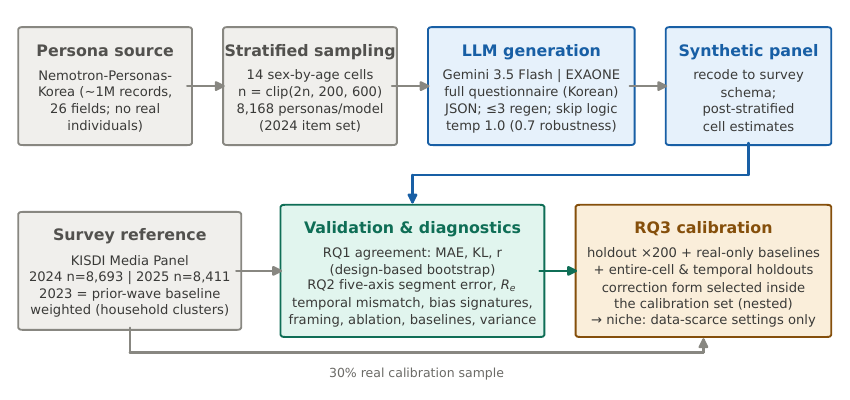}{Study framework: stratified persona sampling and LLM response generation (top row); validation, diagnostics, and signature-matched calibration against the survey reference estimates (bottom row).\label{fig:framework}}

\section{Related Work}
\textbf{LLM-based survey simulation (silicon sampling).} Argyle \emph{et al.}
\cite{argyle} conditioned GPT-3 on demographic backstories drawn from the
American National Election Studies (ANES)
and found that population-level opinion distributions and inter-attitude
correlations matched the real data reasonably well; Aher \emph{et al.} \cite{aher}
and Horton \emph{et al.} \cite{horton} demonstrated simulations of human-subject experiments
and of economic agents, respectively. More recently, generative agents built
from rich individual self-reports were shown to simulate people's survey
responses across tasks, with demographic-only conditioning performing worse than
interview-grounded conditioning \cite{parkjs}, and persona-grounded generation
frameworks have begun to operationalize this practice at scale \cite{polypersona}.
Concurrent work has benchmarked generation methods for closed-ended responses
at scale \cite{ahnert} and validated \emph{individual-grounded} digital
personas (built from real respondents' background variables and survey
histories) against held-out waves of a Dutch probability panel \cite{jia}.
Our study differs in validating a \emph{fully synthetic} persona panel, with no
real individual behind any record, against a Korean national reference survey,
under explicit naive and real-only baselines. Recent work
extends validation beyond English-speaking settings and beyond means:
LLM-generated samples approximated a Chilean probability survey only with
substantial item-level heterogeneity \cite{gonzalez}; a cross-domain benchmark
on the General Social Survey and the World Values Survey found that LLMs
overdetermine demographics, treating identity as more predictive of
attitudes than it is among real respondents \cite{chen26}; and a psychometric
audit against Lithuanian organizational surveys found plausible means but
distorted response structure and acquiescence shifts \cite{lukauskas}. We add
to this line a nationally representative non-English reference and a test of
whether small-sample calibration can close the gap.

\textbf{Validity and bias skepticism.} Santurkar \emph{et al.} \cite{santurkar}
showed that LLM opinion distributions are systematically skewed toward certain
groups; Bisbee \emph{et al.} \cite{bisbee} warned of the perils of replacing
surveys with synthetic data, including variance compression and temporal and
group biases; and Dominguez-Olmedo \emph{et al.} \cite{dominguez} questioned the
robustness of LLM responses. We extend this line of skepticism to a non-English
representative panel.

\textbf{Response-style bias.} Tjuatja \emph{et al.} \cite{tjuatja} showed that
LLMs respond to wording changes differently from humans, and Braun \cite{braun}
reported acquiescence bias. Park \emph{et al.} \cite{park} documented response
homogenization (diminished diversity and near-zero variance). The
acquiescence-consistent level bias we observe in EXAONE and the low-anchoring
tendency in Gemini confirm this family of biases in a Korean survey and
contrast its expression across models.

\textbf{Segment fidelity and fairness metrics.} Although success is often
declared on the basis of aggregate means, demographic conditioning can
redistribute error and undermine subgroup fidelity \cite{taday}. In an audit of
63 persona-based studies, Batzner \emph{et al.} \cite{batzner} found that the
task and target population were often left underspecified and called for their
explicit reporting, a transparency gap our design addresses by documenting both in a public registration. By analogy with the demographic parity difference of the
fairness literature \cite{dwork,agarwal}, we report segment error along five
axes together with a between-group signed-error range and its absolute-error
complements.

\textbf{Bias mitigation and calibration.} Post-stratification and data
augmentation are common approaches to bias mitigation; Wang \emph{et al.}
\cite{wang} statistically combined limited real data with LLM-generated data to
correct bias and reduce estimation error, and Cao \emph{et al.} \cite{cao}
fine-tuned LLMs to match group-level response distributions across countries.
Abels \emph{et al.}~\cite{abels} proposed calibrated value personas, constructed from
sampled value profiles, to improve cross-cultural survey simulation.
In a similar spirit, we show that small-sample holdout calibration cuts the
sex-by-age cell error by about 70\% (a lighter, post-hoc alternative
that requires no model training) while also benchmarking the calibrated panel
against real-only estimators, a comparison prior work omits and one that
bounds the panel's value (Sec.~IV-G).

\textbf{Synthetic persona datasets.} NVIDIA Nemotron-Personas \cite{nemotron} is
a collection of synthetic personas whose demographic attributes (sex, age,
education, occupation, region, and household type) are sampled to match official
national distributions and whose narrative fields (life background, hobbies, and
professional and media orientations) are generated by an LLM conditioned on
those attributes; no record corresponds to a real individual. The Korean
release, which contains about one million records, serves as our conditioning source.
Because the narratives are themselves model-generated, our persona ablation
(Sec.~IV-I) directly tests whether they add information beyond the raw
demographics.

\section{Method}
Fig.~\ref{fig:framework} summarizes the study design: personas drawn by
stratified sampling from Nemotron-Personas-Korea answer the survey
questionnaire through two LLMs; the recoded, post-stratified synthetic
estimates are validated against weighted survey estimates and diagnosed for
structural bias (RQ1--RQ2); and a bias correction learned on a 30\%
calibration subsample, matched to the diagnosed bias signature, is
evaluated on held-out and next-wave real data (RQ3). Each component is
detailed below.

\subsection{Data}
The reference data source is the Korea Media Panel Survey \cite{kisdi}, an
officially approved national statistic (no.\ 405001) fielded annually since 2010 by the
Korea Information Society Development Institute (KISDI) through in-person
household interviews of all household members aged 6 and over, conducted in
June--August of each year. The first wave (2010) covered Seoul, the capital
region, and the six metropolitan cities; the panel has been nationwide (all
17 first-tier divisions) since 2011, and the 2024 wave covered 4{,}006
households and 8{,}693 individuals. The survey panel is a stratified two-stage probability sample. Its frame is
the set of census enumeration districts (2005 Census for the initial panel,
2017 Census for the 2019 supplement), stratified by first-tier division and
urban or rural area type and substratified by apartment or farm-household
share, giving 56 strata. Within strata, enumeration districts (500 for the
initial panel, 100 for the supplement) were selected with probability
proportional to their household counts, and about ten households per district
were enrolled with all members aged 6 and over. The initial 2011 panel (5{,}109 households) had shrunk
to 3{,}556 households by 2019, so a supplementary panel of 1{,}027
households (2{,}436 members) was fielded that year to offset attrition and
panel aging; the integrated panel has been fielded since then. Weights are built
in three steps (the design weight, a within-household nonresponse
adjustment, and raking to population totals by division and sex and by sex
and age band), and separate longitudinal weights additionally adjust for
attrition through a response-propensity model \cite{kisdi}.

We use
individual-level microdata from the 2024 wave (the 2025 wave where relevant,
and the 2023 wave only to construct a prior-wave baseline), taking
person-weighted estimates as the reference. The weight is KISDI's
integrated-panel cross-sectional person weight (variable \texttt{p\_\_wt}), which combines the initial and supplementary panels and is the
weight KISDI recommends for cross-sectional analysis; we use it rather than
the longitudinal weight because each wave serves as a cross-sectional
reference, not as a within-person change measure. Attrition, inherent to a long-running panel, is partly addressed by these weights and by the 2019 refreshment sample; panel conditioning, however, may remain in the observed responses and is not corrected by weighting (limitation (13) in Sec.~VI). Throughout,
we refer to the weighted estimates as the reference (or survey) estimates: they
are high-quality probability-survey estimates, not error-free ground truth, and
carry sampling, recall, and wording error of their own, of which our
design-based intervals (Sec.~III-E) capture only the sampling part. The 2025
wave re-fielded the binary media/AI battery (12 items) but not the constructs,
which were measured only in 2024.

The synthetic persona source is Nemotron-Personas-Korea \cite{nemotron}
(v1.0, released 20 April 2026; about one million records, 26 fields): demographic
attributes are sampled from official Korean statistics by a probabilistic
graphical model, the narrative fields are generated by an open-weight LLM
(Gemma~4~31B) conditioned on those attributes, and the dataset contains adults
only (ages 19 and over). We target digital and AI service-use indicators because they are
policy-relevant, are answerable by both the survey and the conditioned personas,
and, being fast-evolving, provide a stringent test of temporal validity. The
eight binary indicators are past use of generative AI, over-the-top (OTT) video
streaming, YouTube, short-form video, social networking services (SNS), mobile
messenger, metaverse, and paid content subscription; each is a yes/no survey item
recoded to 1/0. The eight 5-point constructs are four consumer-innovativeness
dimensions (functional, hedonic, social, cognitive) and the four core
technology-acceptance dimensions of the unified theory of acceptance and use
of technology (UTAUT) \cite{utaut}, namely performance expectancy, effort
expectancy, social influence, and facilitating conditions, each scored as the mean of its constituent Likert items.

\subsection{Sample design}
Age is banded into seven decade groups (teens through 70s-and-over), yielding 14
sex-by-age cells (persona ages start at 19, so the synthetic teen cell contains
only age-19 personas). Under this stratification, the per-cell sample size was
allocated as $n=\mathrm{clip}(n_{\text{survey}}\times 2,\,200,\,600)$
(a floor and a cap), and personas were drawn uniformly at random with replacement
from each cell's eligible pool (fixed seed). Respondents under 10 fall below the youngest band with a persona counterpart,
the teen band of ages 10--19, whose within-band mismatch is examined in
Sec.~IV-B. They are excluded from all cell-based analyses (RQ2, RQ3, and the
paired comparisons), which use the 8{,}675 respondents aged 10 and over in
2024 and the 8{,}395 in 2025. The YouTube and short-form items are conditional: the survey asks them only of
OTT users, so they were answered by 7{,}346 of those 8{,}675 respondents
(7{,}196 of 8{,}395 in 2025), a set that coincides exactly with the OTT users in
both waves. Both indicators are therefore defined on that subpopulation. We
apply the same gate to the synthetic panels (Sec.~III-D), so cell means,
effective sample sizes, and post-stratification shares use the gated subset for
these two items and the full sample for the other six.
The overall RQ1 reference rates use the full
weighted sample, because restricting them to ages 10 and over shifts indicator
rates by at most 0.7~pp; the weighted under-10 share is 1.1\%.

The real reference samples were $n=8{,}693$ (2024) and $n=8{,}411$ (2025).
For the 2024 item set, the valid synthetic samples were $8{,}168$ (Gemini) and
$8{,}165$ (EXAONE); for the 2025 item set, $7{,}938$ (Gemini) and $7{,}794$
(EXAONE), after excluding responses that still failed format validation after
three regenerations (Gemini 0, EXAONE $\le$1.8\%). One response per persona was
collected for the main analysis (multi-response checks are described below). We
stratified on sex and age because these are the most complete and reliable
persona attributes; the remaining axes (education, employment, region) rely on
approximate crosswalks and are used only for post-hoc segment evaluation.
The constants ($\times 2$, floor 200, cap 600) trade Monte Carlo precision
against cost; a sensitivity analysis that subsamples each cell to caps from 150
to 600 changes the RQ1 mean absolute error (MAE) by at most 0.2~pp (16.8--17.0 for Gemini and
13.7--13.9 for EXAONE; only the sampling noise shrinks with larger caps), so
the specific constants do not drive the findings.

\subsection{Models and rationale}
The two models were chosen to contrast the two classes most relevant to Korean
survey simulation: a frontier closed-weight commercial model
(Gemini~3.5~Flash) as the primary model and an open-weight model developed in
Korea (K-EXAONE-236B-A23B, LG AI Research) as the comparison model. This contrast
lets us ask (i) whether a locale-specialized Korean model reproduces Korean
population responses better than a globally trained commercial model and (ii)
how closed models (accessible only through an application programming
interface, API) and open-weight models differ in reproducibility and
governance: open weights can be pinned and rerun indefinitely, whereas hosted
commercial versions can drift. Both models are cost-efficient at the required
scale and support structured output in JavaScript Object Notation (JSON);
Gemini was accessed through its batch API and EXAONE through a serverless
OpenAI-compatible endpoint. Gemini~3.5~Flash has a documented
knowledge cutoff of January 2025; EXAONE's cutoff is undisclosed. Both hosted
models were served in mid-2026. A serving date after the survey waves does not
by itself imply knowledge of post-cutoff survey outcomes, so temporal claims in
Sec.~IV are framed as consistency rather than attribution. These choices were
documented in the registration on the Open Science Framework (OSF; registration
dfe2z, \url{https://osf.io/dfe2z}). Table~\ref{tab:models} lists the identifiers,
endpoints, and decoding settings of both systems as served.

\begin{table}[!t]
\caption{Model and serving identifiers. Both systems were served in July 2026 (the
randomized-order regeneration of Sec.~IV-K on 3 September 2026); request
payloads (system instruction, item text, decoding parameters) and run logs are
archived in the reproducibility package.}
\label{tab:models}
\centering
\footnotesize
\setlength{\tabcolsep}{3pt}
\begin{tabular}{>{\raggedright\arraybackslash}p{2.0cm}>{\raggedright\arraybackslash}p{2.7cm}>{\raggedright\arraybackslash}p{3.1cm}}
\toprule
 & Primary & Comparison \\
\midrule
Model identifier & \texttt{gemini-3.5-flash} & \texttt{LGAI-EXAONE/}\allowbreak\texttt{K-EXAONE-236B-A23B} \\
Developer & Google & LG AI Research \\
Weights & closed & open (pinnable) \\
Access & Gemini Batch API & FriendliAI serverless, OpenAI-compatible \\
Temperature & 1.0 (main) / 0.7 & 1.0 (main) / 0.7 \\
top\_p & 1.0 & 1.0 \\
Reasoning & thinking level ``low'' & disabled (chat template) \\
Output format & JSON output mode & JSON instruction in prompt \\
Max output tokens & default & 4{,}096 \\
Regeneration & $\le$3 batch rounds & $\le$3 attempts \\
Knowledge cutoff & Jan 2025 (documented) & undisclosed \\
Serving window & Jul 2026; 3 Sep 2026 (Sec.~IV-K) & Jul 2026 \\
\bottomrule
\end{tabular}
\end{table}

\subsection{Generation protocol}
Each Nemotron record was rendered as a persona block comprising a header
(sex, exact age, province, occupation, and education in the source's own
wording) plus the summary, cultural-background, hobbies, and arts/media-orientation
narratives (written in the third person), which carry media-consumption cues relevant to the target indicators;
the system instruction asks the model to answer consistently as that individual
(the value of the narrative is tested by the ablation in Sec.~IV-I). The
primary model was \texttt{gemini-3.5-flash} (thinking=low, top\_p=1.0), and the
comparison model was K-EXAONE-236B-A23B (served via FriendliAI); temperature
was fixed at 1.0 for the main analysis, to reflect natural response
variability, and at 0.7 for a robustness check. EXAONE's thinking was disabled
via the chat template to avoid over-reasoning. All materials (persona
narratives, system instructions, and survey items) were presented in Korean,
using the survey's original item wording and response options. Each persona
answered the full questionnaire in a single call, returning a JSON object that
maps item codes to option numbers. Items were presented to every persona in a fixed order, namely the codebook's
module order (the generative-AI items, the
innovativeness and acceptance battery, then the media-use items), which does
not reproduce the placement of these items within the much longer KISDI
instrument; because all items share one context window, responses may be
subject to order and context effects (preceding items, questionnaire length,
and within-call consistency pressure). We test the order component directly
with a randomized-order regeneration (Sec.~IV-K). Responses with format
violations (missing items, out-of-range or non-integer codes) were regenerated
up to three times. All items were presented to every persona; answers to conditional sub-items were discarded afterwards when the gating item
indicated non-use: YouTube and short-form answers were kept only for personas
reporting OTT use, and generative-AI sub-item answers only for AI users, mirroring the
survey's skip logic in the analysis rather than in the prompt. This excludes
2{,}779 Gemini and 1{,}305 EXAONE personas from the two conditional media
indicators in 2024 (3{,}324 and 1{,}553 in 2025), leaving gated cell sizes of
25--596 (median 478) for Gemini and 402--572 (median 503) for EXAONE in 2024,
and 7--594 (median 364) and 307--548 (median 475) in 2025. Because the gate is
the model's own OTT answer, the gated subpopulation is selected by the model:
for Gemini, which underreports OTT use in the older bands (Appendix
Table~\ref{tab:ageerr-g}), the older gated cells are the personas the model
treats as digitally active. As a sensitivity check, the unconditional
short-form rate, which follows the survey's own skip logic by counting every OTT
non-user as a short-form non-user, is 62.3\% in the survey against 24.8\% (Gemini) and 47.6\%
(EXAONE) post-stratified, so the gap is of the same order on either scale.

Format failures were rare overall (Table~\ref{tab:failures}), but the EXAONE 2025 exclusions were concentrated: 45 of the 144 excluded personas were men in their 40s (45 of 600, 7.5\%, about four times the overall rate). For Gemini, 0.81\% of the 2024
requests (66 of 8{,}168) and 0.06\% of the 2025 requests failed on the first
attempt, all resolved by the second round (no exclusions). The 2024 main run's
batch payload files were overwritten by later runs that reused the same file
names, so its first-attempt counts come from the run log (per sub-batch, not per cell); cell-level first-attempt rates exist for the 2025 run and for the demographic-only 2024 ablation run (highest cell rate 5.9\%). For EXAONE, first-attempt
failures were not logged individually in the live-call setting; the initial
2024 pass left 236 personas (2.9\%) unresolved after three attempts.
Inspection traced these to output truncation at the client's output-token limit (its initial value was not recorded)
rather than to malformed answers; after raising the limit to 4{,}096 tokens,
we regenerated these personas once more under the same three-attempt rule,
leaving 3 (0.04\%) excluded; in 2025, 144 (1.8\%) were excluded, with cell-level
exclusion between 0.2\% and 7.5\%. Because regeneration and exclusion could in
principle select on persona characteristics, we report these rates in full; at the observed levels an exclusion cannot move a cell estimate by more than the excluded share of that cell, i.e., up to 7.5~pp for that cell on the six full-denominator items, up to 8.6~pp on the two OTT-gated items (whose denominator is the 476 gated personas of that cell plus the 45 excluded), and 1.8~pp for the EXAONE 2025 headline under a worst-case assignment of the excluded personas (limitation (14) in Sec.~VI); the three 2024 exclusions are negligible.

Large-scale generation used the Gemini Batch
API (sub-batches of $\le$1{,}000). The 2024 main run comprised 8{,}234 batch requests and the 2025 run 7{,}943.
The archived 2025 payloads average about 1{,}800 input characters, 14.1 million
in total; the 2024 requests, whose payload files were not preserved, carried
the 36-item questionnaire and roughly 4{,}400 characters each. One full panel
per model completed within a day.
Gemini batch generation for this study covered four full panels (the 2024 main
run, the 2025 run, the temperature-0.7 run, and the demographic-only ablation;
52.3 million input and 11.8 million output tokens) and was billed at
KRW~141{,}510 in July 2026, about US\$100, or roughly US\$25 per
8{,}000-persona panel. EXAONE calls in the same month covered six full panels,
including three independent passes and the subsample experiments (about
110{,}000 requests including retries, 220 million tokens), and were billed at
US\$70.44. Results reflect the model snapshots served during the generation
period, July 2026 for the main panels and 3 September 2026 for the
randomized-order regeneration; the reproducibility package preserves the run
logs of every generation run together with the batch request payloads of the
2025 run, the demographic-only 2024 run, and the order experiment.

\begin{table}[!t]
\caption{Format-violation regeneration and exclusion. Gemini (batch): requests
failing the first round were resubmitted in up to two further rounds. EXAONE
(live calls): up to three attempts per persona; first-attempt failures were not
logged individually. Personas = sampled personas per run. The Gemini temperature-0.7 and demographic-only 2024 runs had 51 (0.6\%) and 106 (1.3\%) first-round failures and no exclusions; the EXAONE temperature-0.7 and demographic-only 2024 runs excluded 73 (0.9\%) and 19 (0.2\%) personas. Exclusions by sex-by-age cell (0.2--7.5\%, worst cell men in their 40s, 2025) are given in Sec.~III-D and in the package sheet of cell-level failures.}
\label{tab:failures}
\centering
\footnotesize
\setlength{\tabcolsep}{4pt}
\begin{tabular}{llrrrr}
\toprule
Model & Wave & Personas & 1st-attempt fail & Excluded & Excl.\ (\%) \\
\midrule
Gemini & 2024 & 8{,}168 & 66 (0.81\%) & 0 & 0.00 \\
Gemini & 2025 & 7{,}938 & 5 (0.06\%) & 0 & 0.00 \\
EXAONE & 2024 & 8{,}168 & n/a & 3 & 0.04 \\
EXAONE & 2025 & 7{,}938 & n/a & 144 & 1.81 \\
\bottomrule
\end{tabular}
\end{table}

\subsection{Validation metrics}
RQ1 uses MAE, cosine similarity, Kullback--Leibler (KL) and Jensen--Shannon
(JS) divergences, and item-mean
Pearson correlation between post-stratified synthetic estimates and weighted survey
estimates (computed across the eight indicator pairs without further weighting,
since each rate is already a weighted estimate; Spearman rank correlations are
reported alongside as a robustness check). The post-stratified synthetic estimate is
$q_v=\sum_c w_c\,\bar q_{vc}$, where $\bar q_{vc}$ is the synthetic mean of
indicator $v$ in sex-by-age cell $c$ and $w_c$ is that cell's weighted share of
the real sample. Because binary items (0--1) and constructs (1--5) lie on
different scales, pooling them in a single correlation inflates $r$; we therefore
report item-mean correlations separately for the binary and construct sets.
MAE, cosine, and the divergences are computed over the $V{=}8$ binary
indicators; constructs are compared separately on their native 5-point scale.
Let $p_v$ and $q_v$ ($v=1,\dots,V$) denote the real and synthetic rates
for indicator $v$. Then
$\mathrm{MAE}=\frac{1}{V}\sum_v |q_v-p_v|$,
$\cos=\frac{\mathbf{p}\cdot\mathbf{q}}{\|\mathbf{p}\|\,\|\mathbf{q}\|}$, and the
per-indicator Bernoulli divergence is
$D_{\mathrm{KL}}(p_v\|q_v)=p_v\ln\frac{p_v}{q_v}+(1-p_v)\ln\frac{1-p_v}{1-q_v}$
(averaged over $v$; JS is its symmetrized form). As a between-group bias index
we report the between-group signed-error range $R_e=\max_g e_g-\min_g e_g$,
where $e_g$ is the signed error of group $g$, computed per indicator over the
14 sex-by-age cells and averaged across indicators. Unless another axis is named, group, cell, and segment all refer to a sex-by-age cell. (The index is analogous
to, but distinct from, the demographic parity difference of the fairness
literature \cite{dwork,agarwal}, which applies to model outcomes rather than
to signed errors, so we use the descriptive name.)
Because a range is determined by the two extreme cells and signed errors can
cancel when averaged, we supplement it with the maximum and mean absolute
group error, the between-group standard deviation (SD) of the signed error, and the
worst-estimated group per indicator (Sec.~IV-B), and we interpret all of them
alongside the per-axis MAE rather than in isolation. RQ2 decomposes cell error along five axes (age, sex,
education, employment, and region, defined as Korea's 17 first-tier
administrative divisions).

Uncertainty in the RQ1 MAE is
quantified by a design-based bootstrap ($B=600$): real respondents are
resampled by household cluster, which preserves the within-household
correlation of respondents, and weighted estimates are recomputed in each
replicate, while synthetic personas (independent draws by construction) are
resampled at the row level. (The individual-level file carries 4{,}016
distinct household identifiers, ten more than the 4{,}006 completed household
interviews KISDI reports for 2024; we cluster on the identifier as recorded.) We report 95\% percentile
confidence intervals (CIs). The unequal weights imply a Kish design effect of
2.94 (effective $n\approx2{,}961$), i.e., the variance inflation induced by
weighting; the household resampling adds the clustering of persons within
households. The procedure does not reproduce the survey's stratification or
its first-stage clustering by enumeration district: the microdata identify
the division and area type (the top-level strata) but not the sub-strata or
the districts, so the survey-sampling component of these intervals is
approximate. Comparisons
between two panels evaluated on the same indicators (e.g., Gemini vs.\ EXAONE)
use a \emph{paired} version of this bootstrap. Each replicate draws one
household resample of the survey and one resample of the personas that both
panels answered validly, applies both to the two panels, and records the
difference in MAE, which yields a percentile CI and a two-sided bootstrap
$p$-value for the difference; per-indicator differences are Holm-adjusted
across the eight indicators. Two-sided $p$-values follow the add-one
convention, $p=\min\{1,\,2\min(k_-+1,k_++1)/(B+1)\}$ with $k_\pm$ the number
of replicates on each side of zero, so the smallest attainable value is
$2/(B+1)$ ($B=600$ here, $B=2{,}000$ for the persona-level bootstrap of
Sec.~IV-K). These intervals cover finite-sample survey error and the Monte
Carlo error of the realized synthetic panel; they do not cover generation
stochasticity (the variability of re-querying the same personas), which we
quantify separately with repeated responses and, jointly with the other two
sources, with a hierarchical bootstrap that resamples households, then personas, then
one of each persona's repeated responses (Sec.~IV-J).

Post-stratification aligns only the sex-by-age marginal, which is matched by
design; the comparison therefore isolates the \emph{outcome} distribution given
demographics rather than demographic representativeness, so the reported
agreement is not inflated by demographic matching. Cosine similarity is
reported for completeness but is magnitude-dominated and should not be
overinterpreted; MAE, KL, and item correlation are the primary metrics. Given
$n\approx8{,}000$, almost any difference is statistically detectable; we
therefore emphasize practical magnitude (pp) and report CIs for the headline
claims rather than per-cell significance tests. Two kinds of interval appear
below: design-based 95\% CIs (survey sampling plus Monte Carlo error, as above)
and split-percentile intervals over repeated calibration splits (Sec.~III-F),
which describe stability across splits and are not population CIs. Because the
paper reports many indicator-by-axis comparisons, we do not apply a global
multiplicity correction; instead we label the status of every analysis
(Sec.~III-H, Table~\ref{tab:status}) and report family-level Holm-adjusted results for the principal model comparison, the wording experiment, and the randomized-order experiment (Secs.~IV-A, IV-D, IV-K).

\subsection{Calibration design (RQ3)}
RQ3 was not part of the registered plan, which covered RQ1 and RQ2
(Sec.~III-H); it was formulated after the raw comparison revealed large, structured
error and is therefore a post-hoc analysis that was not preregistered. The 2024 real data were split
within sex-by-age strata into a calibration set (30\%, $n=2{,}598$; fractions
refer to unweighted respondent counts) and a test set (70\%, $n=6{,}077$),
both drawn from the 8{,}675 respondents aged 10 and over. On
the calibration set, the synthetic per-cell bias was learned separately for
each indicator in two forms: (i) a global additive shift, subtracting the
indicator's mean calibration-set bias $\bar b$ from every cell, and (ii) a
linear age regression, $b_c=\beta_0+\beta_1\,\mathrm{age}_c$ with
$\mathrm{age}_c$ the decade-band index of cell $c$, fitted by unweighted least
squares on the 14 calibration cells (so $\beta_1$ is in pp per decade band;
e.g., $\beta_1=-11$ means the bias falls by 11~pp from each band to the next),
with the fitted bias subtracted from each cell. The age form was chosen after
observing the age-structured bias signature in the 2024 panel (per-band 2024 signed errors in Appendix Tables~\ref{tab:ageerr-g} and \ref{tab:ageerr-e}; Sec.~IV-C reports the slopes on the matched-wave 2025 panel), that is, on the same
2024 data rather than before seeing them. To remove this choice from the evaluation
we therefore also implemented a \emph{nested} procedure: within each
calibration set, five inner 50/50 stratified re-splits select, by inner-validation
error, among five forms (a global shift; a linear age term; a quadratic age
term; age plus sex; and age, sex, and their interaction); the selected form is refitted on
the full calibration set and evaluated once on the untouched test set, which
never informs the choice. We report the nested procedure as the principled
estimate, the linear form as the one chosen by inspection (post-hoc), and every candidate
form individually, together with a calibration-cell-size-weighted fit, an
empirical-Bayes partially pooled variant, the fitted
coefficients with their split intervals, residual diagnostics, and the
split-to-split variability of the corrected cell estimates (Appendix
Table~\ref{tab:coef}).

The empirical-Bayes (EB) variant shrinks each calibration cell's observed value
toward the fitted line in proportion to that cell's sampling variance. Writing $b_c$ for the
observed cell quantity, $\hat b_c$ for the fitted value, and
$\sigma^2_c=\hat p_c(1-\hat p_c)/n^{\mathrm{eff}}_c$ for the sampling variance
of the cell mean, with $n^{\mathrm{eff}}_c$ the Kish effective sample size
computed on the respondents who answered that indicator, the shrunk value is
$b^{*}_c=\lambda_c b_c+(1-\lambda_c)\hat b_c$ with
$\lambda_c=\tau^2/(\tau^2+\sigma^2_c)$, where $\tau^2=\max\{0,\widehat{\mathrm{Var}}(b_c-\hat b_c)-\overline{\sigma^2_c}\}$
is the between-cell variance in excess of sampling noise. The between-cell variance, the sampling variances, and the weights are all
estimated inside the calibration set, never on the test cells. Applied to
the synthetic bias it gives the synthetic-target (synthetic-informed) variant (mean $\lambda$
0.51--0.72 across calibration fractions, individual-level splits); applied to the real
cell means with the real-only regression as the fitted line it gives the
real-only shrinkage estimator used as a comparator in Sec.~IV-G (mean
$\lambda$ 0.51--0.69).

The main results use unbounded corrected rates. Clipping the same fitted
predictions to $[0,1]$ lowers the error and is therefore not the conservative
choice: at the 30\% fraction it moves the nested-calibration MAE from 4.92 to
4.46~pp for Gemini and from 4.43 to 4.09~pp for EXAONE (all forms are in the package's calibration-form sensitivity sheet). To account for split
variability, the holdout was repeated over 200 random stratified splits;
percentile intervals over these splits describe stability across repeated
splits (they share observations and are not population CIs). Splits were
individual-level within strata, so members of one household may fall on
different sides of the split; repeating the protocol with household-cluster
splits (whole households assigned together, without stratification) changed
the values by at most 0.1~pp (Sec.~IV-G), ruling out within-household leakage
as a driver.

Because the calibration set is itself a real probability sample, calibration
must be benchmarked against \emph{using that sample directly}. On the same
splits we therefore evaluated three real-only estimators.
(a)~\emph{Direct estimation} is the weighted cell rate in the calibration set;
a cell left empty by the split falls back to the calibration-set grand mean,
and we also report a fallback to the real-only regression.
(b)~The \emph{real-only regression} is an additive regression of the
calibration cell rates on age band and sex, fitted by weighted least squares
with each cell's survey-weight sum as the regression weight, and with
predictions clipped to $[0,1]$. Each estimator is reported in the best
weighting we tested: weighting the synthetic bias fit made it worse
(Sec.~IV-G), whereas the real-only regression is weighted. Clipping is applied to the real-only
regression but not to the corrected synthetic rates, so the reported margins
over the real-only estimators are lower bounds (see the clipping sensitivity above).
(c)~The \emph{calibration-set grand mean} uses the target-wave calibration
data but no cell-specific information, so it is an in-design reference rather
than a zero-data forecast. We swept the calibration
fraction over 1/2/3/5/10/20/30\% (learning curve), recording for every split
the calibration cell sizes, the number of empty cells, which estimator won,
and the paired difference between the calibrated panel and the better of
direct estimation and the real-only regression; and, separately, we held out entire sex-by-age cells (each cell in turn, with 50 repetitions of a 30\% stratified calibration draw from the other 13 cells), calibrating
on the remaining cells only, to test extrapolation to unobserved segments. We
also used a \emph{temporal holdout}: the correction (each form,
including the nested selection carried out within 2024) was learned on the full
2024 data, applied to the 2025 synthetic panel, and evaluated against the 2025
reference estimates. Because the 2025 synthetic responses were generated in
2026, this evaluates the cross-wave transferability of correction coefficients
(a backcast), not forward forecasting.

\subsection{Baselines, ablation, and response variance}
Four additional analyses probe validity. (i) \emph{Naive baselines}: a
grand-mean baseline predicts the overall (weighted) rate for every cell, and a
prior-wave baseline uses the 2023 real cell rates. Because the 2023 wave did
not field the generative-AI and short-form items, the prior-wave baseline
covers six of the eight indicators, and fair comparisons against it are
restricted to this common set. All cell-level comparisons (synthetic,
calibrated, and baseline) are evaluated on the same 14 comparison cells. (ii) \emph{Persona ablation}:
responses are generated for the same personas under an alternative
\emph{demographic-only} conditioning (sex and age only, without the persona
narrative) to test whether the narrative adds value. (iii) \emph{Response
variance}: for a stratified subsample ($\approx$190 personas), $K{=}5$ responses
per persona are collected, and the total variance is decomposed into
within-persona (generation stochasticity) and between-persona (heterogeneity)
components, summarized by the intraclass correlation (ICC),
$\mathrm{ICC}=\sigma^2_{\text{b}}/(\sigma^2_{\text{b}}+\sigma^2_{\text{w}})$.
In addition, the full EXAONE panel was regenerated twice more (three independent
full passes) to test whether generation noise affects aggregate estimates;
agreement was evaluated both per pass and after averaging each persona's three
responses. (iv) \emph{Framing control}: to isolate the effect of the short-form
item's ``OTT'' wording, a stratified subsample (1{,}144 sampled personas per
model, 969--1{,}113 valid) answered the full 2024 questionnaire twice, once with the original
wording and once with a neutral variant that names the short-form venues, with
all else held fixed (Sec.~IV-D).

\subsection{Analysis status and registration}
The analysis plan was registered on OSF as a secondary-data analysis (dfe2z,
registered 3 July 2026; public). The registration was submitted after the
2024 panels had been generated and an initial agreement computation had been
run (2--3 July 2026), and before the calibration, diagnostic, and temporal
analyses reported here were designed; Table~\ref{tab:status} therefore labels
the RQ1--RQ2 plan as registered rather than as strictly preregistered. The registration's own prior-knowledge statement, which states that the synthetic responses had not yet been generated and that no comparison had been computed, was therefore inaccurate when filed; we correct it here and in the package's registration correspondence table. Table~\ref{tab:status} states the status of every
analysis reported below and the deviations from the registered plan, so that
registered, robustness, exploratory, and post-hoc results can be
distinguished. The registered primary agreement measure was the total variation
distance; we report MAE in percentage points instead, because the item set is
dominated by binary indicators for which MAE is directly interpretable and is
the unit of all calibration results (JS divergence and correlations are
reported as registered). The registered mixed-effects regression of cell
agreement on cell size, age group, and item type was not fitted; the
descriptive five-axis decomposition addresses the same question. The
registered chance reference (a uniform or permutation baseline) was replaced by
the stricter real-data baselines of Sec.~III-G. The registered item-type
contrasts (binary vs.\ multi-category items; general digital vs.\ AI items)
were not run as registered: the reported contrast is binary indicators vs.\
5-point constructs, and the three multi-category AI items were not analyzed.
Of the registered robustness summaries, mean absolute differences across
settings are reported and cross-setting correlations are not. A line-by-line
correspondence table accompanies the reproducibility package.

\begin{table}[!t]
\caption{Status of the reported analyses.}
\label{tab:status}
\centering
\scriptsize
\setlength{\tabcolsep}{4pt}
\begin{tabular}{p{2.4cm}p{5.5cm}}
\toprule
Status & Analyses \\
\midrule
Registered after panel generation (RQ1--RQ2) & RQ1 overall agreement; RQ2 error by sex-by-age cell; models, temperatures, stratified allocation, skip logic (generative-AI sub-items), three-attempt regeneration \\
Deviations from registration & MAE replaces total variation distance as the primary measure; five-axis decomposition replaces the registered mixed-effects regression; real-data baselines replace the registered chance reference; the registered item-type contrasts (binary vs.\ multi-category; general digital vs.\ AI) are replaced by the binary-indicator vs.\ construct contrast, and the three multi-category AI items are not analyzed; of the registered robustness summaries, mean absolute differences are reported and cross-setting correlations are not; the EXAONE 2024 run raised the client's output-token limit mid-run and repeated the three-attempt cycle once more for the affected personas (Sec.~III-D); the survey's OTT skip logic for the YouTube and short-form items was applied to the synthetic responses after the registered analyses had first been run (Sec.~III-D); the eight registered age categories are analyzed as seven decade bands, with respondents under 10 (weighted share 1.1\%) excluded from cell-based analyses \\
Prespecified robustness & Temperature 0.7; comparison model \\
Exploratory diagnostics (2024 data) & Bias signatures (age slopes, response style); teen-cell-excluded rerun; 19-year-old comparison; construct correlation structure; sampling-cap sensitivity; agreement-metric supplements (leave-one-item-out, medians, item-bootstrap CIs, Spearman, paired model bootstrap) \\
Exploratory experiments (additional generation runs) & Wording-controlled regeneration; 2025 matched-wave regeneration and reference-year swap; randomized-order regeneration \\
Post-hoc (not preregistered) & RQ3 calibration in its entirety (all correction forms, nested selection, learning curve, entire-cell and temporal holdouts); real-only estimators; naive baselines and ablation; variance and stability analyses \\
\bottomrule
\end{tabular}
\end{table}

\section{Results}
\subsection{RQ1: Overall agreement}
\begin{figure*}[t!]
\centering
\begin{minipage}[t]{0.48\textwidth}\centering\includegraphics[width=\linewidth]{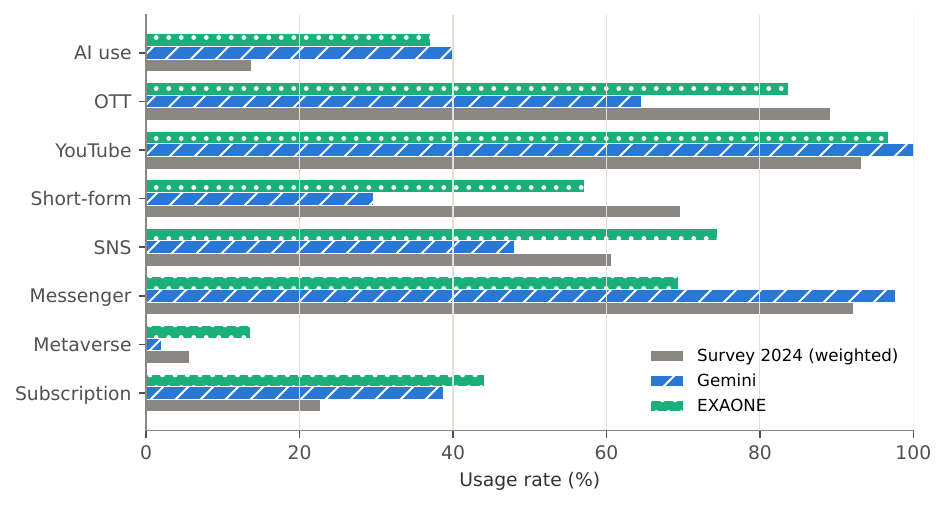}\\{\footnotesize (a)}\end{minipage}\hfill
\begin{minipage}[t]{0.48\textwidth}\centering\includegraphics[width=\linewidth]{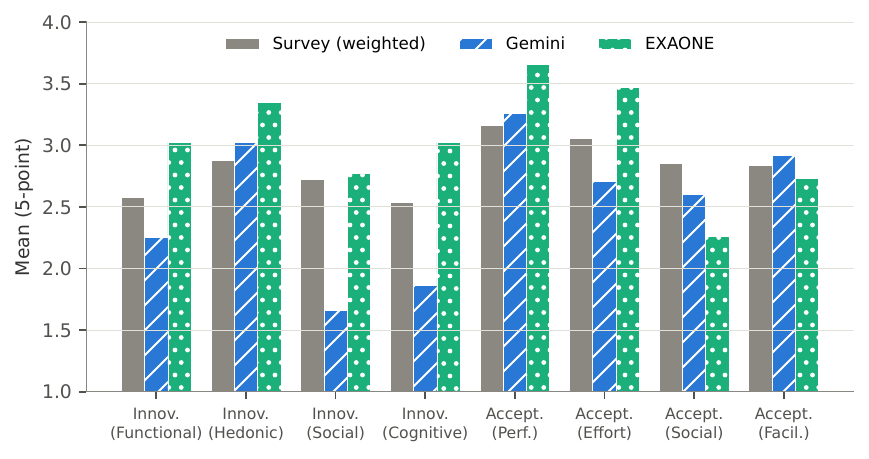}\\{\footnotesize (b)}\end{minipage}
\caption{RQ1 agreement, 2024. (a) Use rates: weighted survey estimates vs.\ Gemini and EXAONE (YouTube and short-form among OTT users, Sec.~III-B); design-based sampling standard errors are below 1~pp per indicator (95\% half-widths at most about 1.8~pp; Sec.~III-E), smaller than the bar resolution. (b) Construct means (5-point): survey vs.\ synthetic.\label{fig:rq1}\label{fig:constructs}}
\end{figure*}

Against the 2024 reference estimates, the overall MAE was 17.0~pp (Gemini;
design-based 95\% CI [16.5, 17.4]) and 13.8~pp (EXAONE; [13.3, 14.4]).
A paired design-based bootstrap of the difference, computed on the
age-10-and-over sample and on the 8{,}165 personas that both models answered
validly, gave 3.2~pp (95\% CI [2.5, 3.8]; two-sided bootstrap $p=0.003$, the
smallest value attainable with $B=600$), so EXAONE was closer to the survey
distribution. The advantage was not uniform across indicators. After Holm
adjustment, EXAONE was closer on generative-AI, OTT, YouTube, and short-form use
(differences of 2.8, 19.2, 3.4, and 27.5~pp) and Gemini on messenger, metaverse,
and subscription use (18.1, 4.6, and 5.3~pp); only the SNS difference was
not significant after adjustment (0.3~pp, adjusted $p=0.88$).
For comparison, the design-adjusted sampling standard error (Sec.~III-E) is
below 1~pp per indicator (95\% half-width at most 1.8~pp), an order of
magnitude smaller than the observed MAE, so the gap reflects bias, not noise. Cosine similarity was high (0.949/0.971);
KL divergence (0.189/0.066) and the binary item-mean correlation (Pearson
0.811/0.912; Spearman 0.83/0.88)
consistently favored EXAONE (Table~\ref{tab:rq1}, Fig.~\ref{fig:rq1}(a)); the
largest between-model differences were on short-form, OTT, and messenger use
(Gemini worse on the first two, EXAONE on the third). Computed over only eight items, these correlations carry wide uncertainty
(item-bootstrap 95\% CIs [0.48, 0.97] and [0.72, 0.99]) and are descriptive. No single item drives the 2024 ranking of the two models:
the median absolute error was 14.4/13.1~pp, and leave-one-item-out MAE stayed
within 13.7--18.9~pp (Gemini) and 12.5--15.3~pp (EXAONE), with Gemini above
EXAONE throughout. The moderate-to-high correlations with a large MAE indicate that the models
ranked behaviors roughly correctly but misestimated their levels. The innovativeness and UTAUT-based
acceptance constructs (5-point) showed a mean-level MAE of 0.373 (Gemini) and
0.380 (EXAONE), about 9\% of the 4-point scale range, with opposite signs
(Gemini underestimated, EXAONE overestimated; see the response-style analysis
below). Neither model can be taken as an off-the-shelf substitute for measured
acceptance attitudes. EXAONE's construct-level correlation was weak
($r=0.500$ vs.\ Gemini's 0.788; Spearman 0.52 vs.\ 0.81), so its advantage on
binary indicators did not extend to construct ordering
(Fig.~\ref{fig:constructs}(b)). Agreement in construct means does not establish
psychometric equivalence, and we do not claim it: within-panel inter-construct
correlations (28 pairs) differed from the survey's by 0.085 (Gemini) and 0.108
(EXAONE) in mean absolute terms, the correlation between the synthetic and
survey correlation vectors was 0.01 (Gemini) and 0.35 (EXAONE), and construct
standard deviations were compressed to 71\% (Gemini) and 85\% (EXAONE) of the
survey's (Appendix Table~\ref{tab:concorr}). Our construct-level claims are
therefore restricted to agreement in means; factor structure and measurement
invariance are outside the scope of this study.

\begin{table}[!t]
\caption{RQ1 overall agreement metrics. Ref.: reference-survey year. The 2024 rows use
the 2024-item panel; the 2025 rows use the matched-wave (2025-item) regeneration.
Correlations are reported separately for the 8 binary items ($r_{\text{bin}}$)
and the 8 constructs ($r_{\text{con}}$; not measured in 2025), because pooling
the two scales inflates the correlation. MAE 95\% CIs (design-based household-cluster bootstrap,
$B=600$): Gemini 2024 [16.5, 17.4], EXAONE 2024 [13.3, 14.4].}
\label{tab:rq1}
\centering
\setlength{\tabcolsep}{3pt}
\begin{tabular}{llcccccc}
\toprule
Ref. & Model & MAE (pp) & Cosine & KL & JS & $r_{\text{bin}}$ & $r_{\text{con}}$ \\
\midrule
2024 & Gemini & 17.0 & 0.949 & 0.189 & 0.029 & 0.811 & 0.788 \\
2024 & EXAONE & 13.8 & 0.971 & 0.066 & 0.018 & 0.912 & 0.500 \\
2025 & Gemini & 19.4 & 0.914 & 0.235 & 0.052 & 0.703 & -- \\
2025 & EXAONE & 15.2 & 0.953 & 0.114 & 0.030 & 0.787 & -- \\
\bottomrule
\end{tabular}
\end{table}

\subsection{RQ2: Segment error}
The five-axis segment MAE was 15.2--18.3~pp (Gemini) and 13.9--15.6~pp (EXAONE),
with Gemini performing worst on the age and education axes (18.3 and 17.3~pp) and EXAONE trailing Gemini only on the region axis (15.6 vs.\ 15.4~pp). The combined sex-by-age cell
MAE, the reference quantity for the calibration analysis below, was 18.3~pp
(Gemini) and 15.1~pp (EXAONE). The between-group signed-error range $R_e$ was
49.6~pp (Gemini) and 34.7~pp (EXAONE): for Gemini, the signed error on a typical
indicator spanned nearly 50~pp between the most overestimated and the most
underestimated demographic cells, a spread large enough to flip the sign of
segment-level conclusions. The absolute-error complements lead to the same conclusion
with less dependence on the two extreme cells: the maximum absolute group
error averaged 42.5/29.7~pp, the mean absolute group error (which coincides with the combined cell MAE) 18.3/15.1~pp, and
the between-group SD of the signed error 16.8/10.6~pp (Table~\ref{tab:rq2}).
The worst-estimated group per indicator was, for Gemini, teenage women on
generative-AI use ($+71$~pp) and men in their 40s on short-form ($-65$~pp) and
SNS use ($-45$~pp); for EXAONE, teenage men on generative-AI use ($+47$~pp)
and women aged 70 and over on SNS use ($+42$~pp).

Because the synthetic teen cell contains only age-19 personas whereas the
survey's teen band spans ages 10--19 (Sec.~III-B), the two teen cells are not
population-comparable. We keep the 14 cells as the headline unit because they are the unit of the calibration and entire-cell holdout analyses, and report the 12-cell results that exclude both teen cells as a comparability sensitivity analysis alongside them (Appendix Table~\ref{tab:teenexcl}), qualifying age-segment conclusions by both. On the
12 cells, $R_e$ was 39.5/31.2~pp, the mean absolute group error 17.6/15.0~pp,
and the between-group SD 14.2/10.1~pp; Gemini's negative age slopes on the core stereotype indicators (OTT and generative-AI use) persist, the endpoints of the five-axis MAE range move by at most 0.8~pp (individual axes by up to 1.6~pp), and most calibration and baseline orderings are preserved. The exception is Gemini's entire-cell holdout: nested calibration beats the real-only regression on the 14 cells (6.2 vs.\ 8.6~pp) but not on the 12 non-teen cells (6.7 vs.\ 6.2~pp). Overall RQ1 MAE rises without the teen cell
(17.0$\to$17.7~pp for Gemini): the teen cells inflated the range statistic but
not the headline error levels. A like-for-like check confirms
that part of the teen-cell error is a band artifact: when the survey's teen
cell is restricted to its 75 respondents aged 19, the teen-cell MAE falls from
22.7 to 12.6~pp (Gemini) and from 16.1 to 11.9~pp (EXAONE), so roughly half (Gemini) or a quarter (EXAONE) of
the apparent teen error reflects comparing 19-year-old personas with
respondents aged 10--18 rather than model error (75 respondents, no interval). Gemini's error is concentrated in particular indicator-by-band combinations rather than in one age region: it deepens through the 40s to 60s bands on OTT, short-form, and SNS use (OTT $-64$~pp at 60--69), while averaged over the eight indicators the 70-and-over band carries the smallest error of the seven (11.5 vs.\ 20.5~pp in the 40s, from Appendix Table~\ref{tab:ageerr-g}; the seven bands average to the 18.3~pp of Table~\ref{tab:rq2}).

\begin{table}[!t]
\caption{RQ2 five-axis segment MAE (pp), combined sex-by-age cell MAE, the
between-group signed-error range ($R_e$), and its absolute-error complements. The five-axis rows are computed over each axis's own groups; the lower block is over the 14 sex-by-age cells. All values are averaged across the eight indicators.}
\label{tab:rq2}
\centering
\begin{tabular}{lcc}
\toprule
Axis & Gemini & EXAONE \\
\midrule
Age & 18.3 & 15.1 \\
Sex & 15.4 & 14.0 \\
Education & 17.3 & 14.7 \\
Employment & 15.2 & 13.9 \\
Region (17 divisions) & 15.4 & 15.6 \\
\midrule
Sex-by-age cells (combined) & 18.3 & 15.1 \\
Signed-error range $R_e$ & 49.6 & 34.7 \\
Max.\ absolute group error & 42.5 & 29.7 \\
Between-group SD of signed error & 16.8 & 10.6 \\
\bottomrule
\end{tabular}
\end{table}

\subsection{Structural bias signatures}
\begin{figure*}[t!]
\centering
\begin{minipage}[t]{0.48\textwidth}\centering\includegraphics[width=\linewidth]{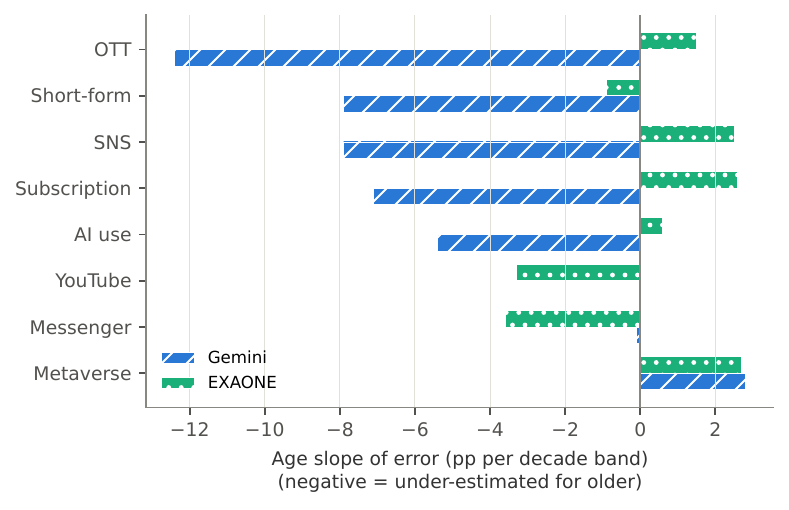}\\{\footnotesize (a)}\end{minipage}\hfill
\begin{minipage}[t]{0.48\textwidth}\centering\includegraphics[width=\linewidth]{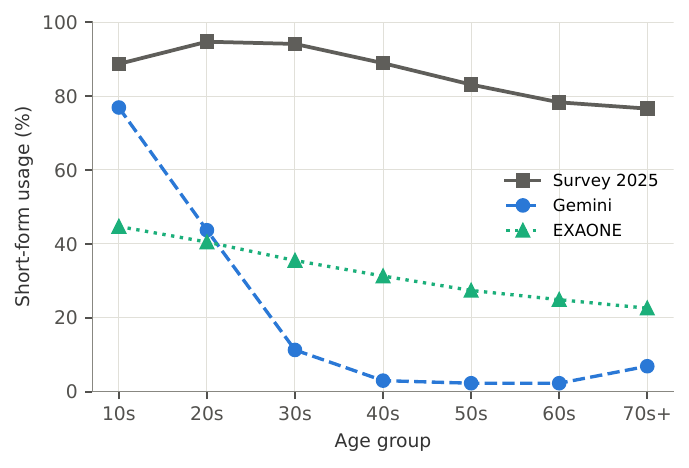}\\{\footnotesize (b)}\end{minipage}
\caption{Age structure of the error on the matched-wave 2025 panel. (a) Age slope of the signed error per indicator (seven pooled age bands; negative = older groups underestimated). (b) Short-form usage by age: gently declining survey trend vs.\ steep synthetic decline.\label{fig:slope}\label{fig:shortform}}
\end{figure*}

An age-slope analysis of the signed error (synthetic $-$ survey), fitted
across the seven pooled age bands on the matched-wave 2025 panel
(Sec.~IV-E), revealed steep negative slopes for Gemini across most digital
indicators (OTT $-12.4$, short-form $-7.9$, SNS $-7.9$, subscription $-7.1$~pp per
decade band; a slope of $-12.4$ means that the signed error falls by 12.4~pp
from each band to the next, e.g., from the 40s to the 50s), consistent with an amplified ``digital = young'' stereotype (Fig.~\ref{fig:slope}(a)). The 2024 panel shows the same signature: seven-band slopes computed from the per-band signed errors in Appendix Table~\ref{tab:ageerr-g} are about $-11$ (generative AI), $-10$ (OTT), $-9$ (subscription), $-7$ (SNS), $-6$ (short-form), and $-1$ (YouTube)~pp per band for Gemini, with messenger and metaverse use running the other way ($+2$ each), versus roughly $-5$ to $+4$ for EXAONE (Appendix Table~\ref{tab:ageerr-e}); the calibration-split 2024 slopes are in Appendix Table~\ref{tab:coef}. Digital use among older Korean adults is high: short-form use remained
between about 77\% and 95\% across all age bands in the 2025 wave
(Fig.~\ref{fig:shortform}(b)), a pattern that the models failed to reproduce. Gemini's 5-point responses were also
anchored low: its mean across the eight post-stratified construct estimates
was 2.54 against 2.83 in the weighted survey, and fewer than 0.5\% of its raw
item responses used the top point. EXAONE showed flatter slopes but a positive
level bias: its construct mean was 3.03, and its mean post-stratified rate
across the eight binary indicators was 59.5\% against the survey's 55.9\%,
whereas Gemini's was 52.5\%.

\subsection{Item framing}
The survey's short-form item frames the behavior as a ``short-form \emph{OTT
service}.'' In Korean usage, ``OTT'' connotes subscription video
streaming (Netflix, TVING), so this phrasing steers a literal-minded respondent
away from counting YouTube Shorts, Instagram Reels, or TikTok, the dominant
short-form venues, whereas the YouTube item names Shorts. The models
behaved like such a literal-minded respondent (2025 wave): among personas
routed into the module, YouTube use was near-universal (Gemini 100.0\%, EXAONE
92.4\%, survey 95.5\%) while short-form use collapsed to 27.1\% and 32.4\%
against 86.5\% in the survey. This inconsistency is not an artifact of the shorter 2025
item set. In the 2024 run, whose 36-item questionnaire matches the full item
set used in this study, short-form use is already underestimated (Gemini
42.5\%, EXAONE 57.6\% vs.\ survey 69.6\%), and the same within-run
inconsistency holds: among synthetic YouTube users only 42.5\% and 58.6\% report
short-form use versus 73.1\% in the survey. Synthetic rates here are unweighted
diagnostics on the gated subsample against weighted survey benchmarks; post-stratified, 2024
short-form use is 29.6\% and 57.1\% against 69.6\%. The framing effect is therefore present under length-matched
conditions; the 2025 regeneration only deepens it.

To isolate the word-level
cause, we ran the wording-controlled regeneration of Sec.~III-G on a stratified subsample
(1{,}144 sampled personas), holding the personas, model, temperature, and the
other 35 items fixed (only the short-form and YouTube answers were retained, so the OTT skip logic cannot be applied; both arms are unconditional and the experiment identifies the contrast between arms, not the level) and changing only the short-form item's
wording, removing ``OTT'' and naming the venues (YouTube Shorts, Instagram Reels,
TikTok), as the YouTube item already does. This wording change moved
short-form use by tens of percentage points: Gemini 31.6$\to$76.8\% and EXAONE
52.9$\to$81.3\%. Both arm rates run over all personas, so they are not on the same footing as the gated survey rate of 69.6\%, and no human arm was fielded. Each arm used a single generation call per persona without the regeneration loop of Sec.~III-D, so first-attempt failures were dropped (2.7--2.8\% of personas for Gemini, 12.8--15.3\% for EXAONE). Under the most adverse and the most favorable assignments of every dropped persona (counting all excluded treatment personas as non-users and all excluded control personas as users, or the reverse), with both arms expressed over the 1{,}144 personas assigned to each, the change lies between 41.2 and 46.7~pp for Gemini and between 10.8 and 38.9~pp for EXAONE (a deterministic bound on the exclusions, not a confidence interval; package framing-test sheet). The Gemini contrast is therefore robust to any assignment of its dropped personas, whereas the EXAONE effect could be as small as about 11~pp, so its magnitude should be interpreted with caution and the quantitative claim rests on Gemini. Because per-persona pairing was not retained, the two arms (969--1{,}113 valid personas each) are analyzed as independent samples: the change is 45.2~pp [41.5, 48.9] for Gemini and 28.3~pp [24.3, 32.3] for EXAONE (two-proportion $z$-tests, $p<10^{-10}$, Holm-adjusted across the two contrasts); the conditional rates below are descriptive because their denominators were not retained. The within-run
conditional rate likewise moved (P(short-form\,$\mid$\,YouTube): Gemini 32.7$\to$79.6\%, EXAONE 59.9$\to$92.6\%; survey 73.1\%, on the gated scale). The independent-sample intervals above would be conservative if the within-persona correlation were nonnegative, an assumption that cannot be verified because the pairing was not retained. Framing alone is therefore sufficient to move the aggregate rate by tens of percentage points, which is inconsistent with a wording-independent stereotype at the aggregate level (the age profile under the neutral wording was not examined); how much of the fixed-order collapse framing accounts for, and how the models' reaction compares with a human respondent's, cannot be identified here, because the wording and order experiments are on different scales and no human arm was fielded. Table~\ref{tab:framing} summarizes the experiment; the age
profile of the short-form error under the original wording appears in Fig.~\ref{fig:shortform}(b).

\begin{table}[!t]
\caption{Wording-controlled regeneration (1{,}144 sampled personas per model;
valid control/treatment responses 1{,}113/1{,}112 for Gemini and 969/998 for
EXAONE; the same personas answered both arms, analyzed as independent samples; one call per persona without regeneration; all else fixed). Changing only the short-form item's wording (removing
``OTT'' and naming the venues) moves short-form use by 45.2~pp (Gemini) and 28.3~pp (EXAONE) between the two arms.
Survey rates are 2024 weighted estimates among OTT users; the synthetic arms are unconditional (Sec.~IV-D), so the survey column is a reference level, not a like-for-like comparison.
95\% CIs on the change exclude zero.}
\label{tab:framing}
\centering
\setlength{\tabcolsep}{4pt}
\begin{tabular}{llccc}
\toprule
Model & Metric & Original & Neutral & Survey \\
\midrule
Gemini & Short-form use & 31.6 & 76.8 & 69.6 \\
Gemini & P(short-form\,$\mid$\,YouTube) & 32.7 & 79.6 & 73.1 \\
EXAONE & Short-form use & 52.9 & 81.3 & 69.6 \\
EXAONE & P(short-form\,$\mid$\,YouTube) & 59.9 & 92.6 & 73.1 \\
\bottomrule
\end{tabular}
\end{table}

\subsection{Temporal misalignment}
\begin{figure*}[t!]
\centering
\begin{minipage}[t]{0.48\textwidth}\centering\includegraphics[width=\linewidth]{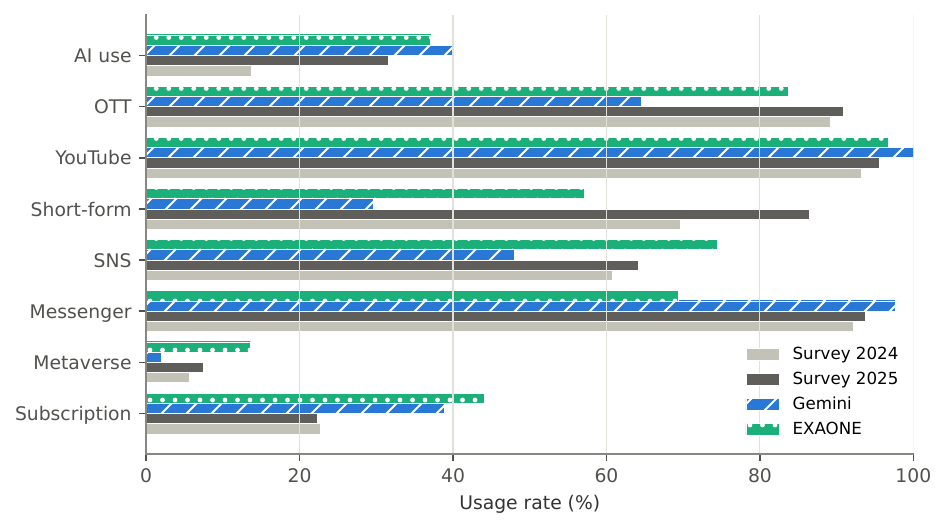}\\{\footnotesize (a)}\end{minipage}\hfill
\begin{minipage}[t]{0.48\textwidth}\centering\includegraphics[width=\linewidth]{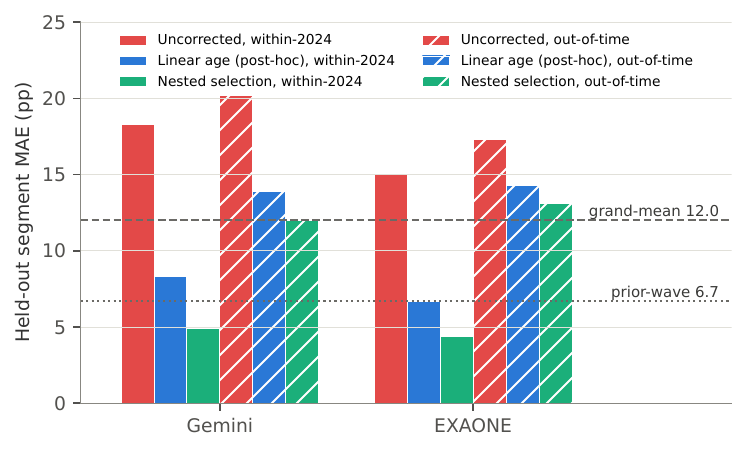}\\{\footnotesize (b)}\end{minipage}
\caption{Temporal comparisons. (a) Reference-year comparison: survey 2024/2025 vs.\ the 2024-item synthetic panels. (b) Held-out sex-by-age cell MAE for the uncorrected panel, the post-hoc linear age correction, and the nested selection: within-2024 (contemporaneous) vs.\ out-of-time (learned on 2024, tested on 2025). Dashed/dotted lines mark the target-year (2025) grand-mean (12.0) and prior-wave (6.7) references. Calibration cuts the error within-wave, but out-of-time the calibrated panel stays at or above both references (Sec.~IV-G).\label{fig:temporal}\label{fig:calib}}
\end{figure*}

In the survey, generative-AI use rose from 13.7\% (2024) to 31.6\% (2025),
and short-form use rose from 69.6\% to 86.5\%. We examined the resulting temporal
mismatch in two ways (all gaps post-stratified, with the cell weights taken from the reference wave). (a) \emph{Reference-year
swap}: evaluating the 2024-item synthetic panel against the 2025 reference estimates
shrank the AI gap from $+26.2$ to $+7.3$~pp (Gemini) and from $+23.4$ to
$+4.9$~pp (EXAONE), so the synthetic estimates sat near the 2025 adoption
level; yet the same swap \emph{widened} the short-form gap
($-40.0\to-58.0$~pp for Gemini; $-12.5\to-29.8$ for EXAONE), a pattern consistent
with a framing-linked rather than temporal short-form error; the
overall MAE barely moved (17.0$\to$17.5~pp for Gemini). (b) \emph{Matched-wave
regeneration}: regenerating with the 2025 item set and comparing against the
2025 reference kept the AI gap small ($+5.3$/$+11.7$~pp) but did \emph{not}
improve the overall MAE (19.4/15.2~pp; paired model difference 4.3~pp [3.6, 4.9] on the 7{,}794 common personas), because short-form use collapsed
($-69.4$/$-54.5$~pp; survey 86.5\% vs.\ Gemini 17.1\%); the drop in the
binary item-mean correlation (0.811$\to$0.703 for Gemini) was likewise driven by this
collapse. This single framing-sensitive item shapes the 2025 headline: excluding
the short-form item, the 2025 MAE would be 12.3/9.5~pp (median absolute
error over the remaining seven items 5.7/11.1~pp), better than the 2024 level on the same seven items (13.7/14.0~pp), so the ``no improvement under
alignment'' result is carried largely by the framing failure diagnosed in
Sec.~IV-D rather than by broad temporal misfit. Aligning the reference year thus
narrowed the gap on the fast-moving new-technology indicator but did not restore overall
validity as measured by the full item set (Fig.~\ref{fig:temporal}(a)).
These patterns are \emph{consistent with} temporal misalignment, but the design
cannot isolate its source. In time order, the 2024 fieldwork preceded Gemini's documented cutoff (January 2025), which preceded the 2025 fieldwork, the persona release (v1.0, April 2026), the main generation runs (July 2026), and the order experiment (September 2026). Three confounded channels could produce these patterns: (i) the response model's pretraining knowledge (Gemini's
documented cutoff, January 2025, postdates the 2024 fieldwork and predates the
2025 wave; EXAONE's cutoff is undisclosed); (ii) the persona source, whose
narratives were generated in early 2026 by an LLM from official statistics
(Sec.~III-A) and may embed 2025-level media habits; and (iii) generic diffusion
reasoning, i.e., a model extrapolating that a new technology is more widely
used than the survey year shows. Channel (ii) affects both response models
identically, and the overestimation appears for both. Because EXAONE's cutoff
is undisclosed, however, that shared pattern does not by itself discriminate
among the three channels: the design cannot separate them, and we therefore do
not attribute the overestimation to any one of them.

\subsection{Robustness}
The mean absolute difference in post-stratified rates between temperatures 1.0
and 0.7 was 0.5~pp for Gemini and 3.7~pp for EXAONE. A paired design-based
bootstrap of the overall MAE difference between the two temperatures gives
0.1~pp (95\% CI [$-0.1$, 0.3]) for Gemini and 2.1~pp ([1.7, 2.5]) for EXAONE.
The lower temperature left Gemini's overall MAE essentially unchanged
(17.0 vs.\ 16.9~pp) but reduced EXAONE's from 13.8 to 11.7~pp on the
age-10-and-over sample, mostly on OTT, short-form, messenger, and metaverse use
(YouTube and SNS use worsened by 2.2 and 3.5~pp). The main analysis retains the registered
temperature of 1.0; the EXAONE headline level should therefore be read as
temperature-sensitive.

\subsection{RQ3: Calibration}
Across 200 repeated stratified holdouts, the uncorrected sex-by-age cell MAE
(Gemini 18.3, EXAONE 15.1~pp) fell to 14.2/8.4~pp with a global additive
correction, to 8.3/6.7~pp with the post-hoc linear age correction, and to 4.9/4.4~pp with the nested procedure, in which the
correction form is selected inside the calibration set (Table~\ref{tab:rq3},
Fig.~\ref{fig:calib}(b)). The nested procedure chose a quadratic age term in
66\%/68\% of indicator-by-split cases (linear 14\%/15\%, sex terms 20\%/17\%);
a quadratic term fitted directly gave 5.0/4.5~pp, so the nested estimate is in
effect the quadratic one, and the linear form chosen by inspection understated
the attainable correction. Weighting the linear fit by calibration-cell
effective sample size did not improve it (8.7/7.0~pp), and adding sex or an
age-by-sex interaction changed it by at most 0.2~pp. The split-percentile
interval of the nested MAE was [4.5, 5.4]/[4.0, 4.9]~pp, and household-cluster
splits changed the values by at most 0.1~pp (linear 8.4/6.7, nested
5.0/4.5~pp), which rules out within-household leakage. Appendix Table~\ref{tab:coef}
lists the fitted intercepts and slopes per indicator with their split
intervals, residual diagnostics, and the split-to-split half-width of the
corrected cell estimates (1.7--3.5~pp). An empirical-Bayes partially pooled
correction, which shrinks each calibration cell's observed bias toward the
fitted age line in proportion to the cell's sampling variance (mean weight on
the observed cell 0.51--0.72 across fractions), reaches 3.3/3.5~pp at the
30\% fraction, below direct estimation (3.6~pp), because the corrected estimate
becomes a precision-weighted combination of the real cell mean and the
line-corrected synthetic value. From the 10\% fraction upward it is the most
accurate synthetic-using estimator we tried, for both models and both split
types. At the 5\% fraction the ranking is model-dependent (EB 5.7 against
6.0~pp for a directly fitted quadratic under Gemini; 5.6 against 5.5~pp under
EXAONE), and below that fraction the quadratic is the more accurate form for
both models. The real-only counterpart of the EB variant is evaluated below.

The contribution of each correction term tracked the model's
bias signature: for EXAONE, whose bias is dominated by per-indicator offsets, the global
correction alone captured most of the linear gain (15.1$\to$8.4, vs.\ 6.7
with the age term added), whereas for Gemini, whose bias is age-structured,
the global shift helped little (18.3$\to$14.2) and the age terms were
essential.

\textbf{Calibration does not beat using the real subsample directly beyond a
few hundred responses.} The calibration set is itself a real probability
sample, so the decisive comparison is against real-only estimators
evaluated on the same splits (Table~\ref{tab:realonly}). At the 30\% fraction
($n=2{,}598$), direct weighted estimation from the calibration set alone
achieved 3.6~pp, better than the nested-calibrated panel in every one of the
200 splits, by 1.3~pp [0.7, 1.9] for Gemini and 0.8~pp [0.3, 1.4] for EXAONE,
and better than the linear form by 4.7~pp [4.0, 5.3] and 3.0~pp [2.3, 3.6].
Every paired difference in this section is reported as the error of the losing estimator
minus that of the winning one, with the split-percentile interval in the same
orientation. The real-only regression
achieved 7.1~pp. The calibration-size sweep locates the crossover
(Table~\ref{tab:realonly}). With the nested correction, the calibrated panel
beat the better of direct estimation and the real-only regression in 76--96\%
of splits (Gemini) and 94--99\% (EXAONE) from 1\% to 5\% of the sample ($n\approx80$ to 430; 9.3/8.5
vs.\ 10.1~pp at 1\%, 6.1/5.8 vs.\ 7.3~pp at 5\%). At 10\% ($n\approx860$)
direct estimation and the calibrated Gemini panel tied (5.5~pp each, the panel
ahead in 50\% of splits) while the calibrated EXAONE panel still led (5.1~pp),
and from 20\% upward ($n\ge1{,}730$) direct estimation dominated both
(4.2 vs.\ 5.1/4.6~pp). Table~\ref{tab:realonly}
gives the win rates at every fraction. Ranking all four estimators split by split, the nested-calibrated panel was the most accurate most often up to 3\% (Gemini 60--71\%, EXAONE 71--94\% of splits) and, for EXAONE, at 5\% (58\%); the real-only shrinkage estimator led from 5\% (Gemini) and 10\% (EXAONE) upward (63--85\%); direct estimation was never first below 10\% and first in 26--33\% of splits at 20--30\%; the real-only regression was first in at most 21\% (Gemini at 1\%; per-fraction values in the package). Direct estimation first won at the tested size
$n\approx1{,}730$ for both models. Throughout, a crossover is the bracket between the last tested size the
calibrated panel won and the first it lost, so against direct estimation it is
roughly 430--1{,}730 real responses for Gemini, whose 10\% point is a tie, and
860--1{,}730 for EXAONE, about 30--125 respondents per cell. The workflow of Sec.~V uses the first tested size at which the calibrated panel stopped winning (860 for Gemini, a tie; 1{,}730 for EXAONE). Against direct estimation the linear form alone crosses over earlier, between $n\approx170$ and 250 for Gemini and between $n\approx430$ and 860 for EXAONE; against the real-only regression Gemini's linear form never wins at any tested size, whereas EXAONE's beats it throughout. Because the
stratified splits allocate at least two respondents to every cell even at 1\%
(median six), no calibration cell was empty under that design; for the two OTT-gated items the smallest calibration cell held about two respondents at 1\% and was empty in 1\% of splits, in which case the item is fitted on the remaining cells; under nonstratified
household splits at 1\% ($n\approx86$), 24\% of splits contained at least one
empty cell (3\% at 2\%, 0.5\% at 3\%, none from 5\% upward), and replacing the empty-cell
fallback (grand mean) by the real-only regression prediction changed direct
estimation by at most 0.03~pp, so the fallback rule does not drive the
comparison.

A partially pooled real-only estimator narrows the niche further.
Shrinking the calibration cell means toward the real-only regression line (the
empirical-Bayes device above, with no synthetic data) beats direct estimation
at every fraction (Table~\ref{tab:realonly}, EB column). Against it, on the
same splits, the nested-calibrated panel is ahead by 1.4/2.3~pp ([$-1.2$, 4.0] and
[0.03, 4.62]; Gemini/EXAONE) at 1\% ($n\approx80$), by 0.4/0.8~pp at 2\%
($n\approx170$) and by 0.1/0.4~pp at 3\% ($n\approx250$); for EXAONE it is
also ahead by 0.1~pp at 5\%. The real-only estimator wins from
$n\approx430$ for Gemini and from $n\approx860$ for EXAONE, so the shrinkage
crossover is roughly 250--430 and 430--860 responses. The synthetic-target
version of the same estimator also beats direct estimation at every fraction,
but improves on its real-only counterpart by at most 0.7~pp (EXAONE at 1\%)
and by 0.3~pp or less for Gemini (per-fraction values in the package). That
bound applies within the shrinkage family only; the nested-calibrated panel is
not a member of it and carries the 1.4 and 2.3~pp margins reported above. The
Gemini interval includes zero, so its margin is not resolved by the
split-to-split variation, whereas the EXAONE interval, [0.03, 4.62]~pp at two
decimals, just excludes it; either way these intervals describe the stability
of overlapping resamples, not population uncertainty or a significance test (Sec.~III-F).

When entire sex-by-age cells were held out (no real data for the
target cell entering calibration), the nested-calibrated panel extrapolated
better than the real-only regression for both models on the 14 cells
(6.2/5.7 vs.\ 8.6~pp; under the linear form, 9.6/7.7~pp, so only EXAONE beat
the regression), but on the 12 non-teen cells only EXAONE retained the advantage
(4.8 vs.\ 6.1~pp; Gemini 6.7 vs.\ 6.2~pp; Appendix Table~\ref{tab:teenexcl}). The synthetic panel's operational niche is therefore narrow: calibration
samples below roughly 250--860 responses against a real-only shrinkage
estimator and below roughly 430--1{,}730 against direct estimation, depending
on the model, together with unobserved segments. Against the better of direct
estimation and the real-only regression, the mean paired margin exceeds 1~pp for EXAONE at every fraction from 1 to 5\% (1.4--1.7~pp) and for Gemini from 2 to 5\% (1.1--1.3~pp, peaking at 3\%); against the
shrinkage estimator it does so only at the smallest calibration size
(1.4/2.3~pp at $n\approx80$). For the comparison against the better of the two real-only estimators, the winning comparator is
chosen split by split, so the mean paired margin is up to 0.1~pp smaller than
the difference of the column means in Table~\ref{tab:realonly}; against the
shrinkage estimator, which is a single fixed comparator, the two differ only
by rounding.

Under the stricter \emph{temporal holdout} (correction learned on 2024, applied
to the 2025 panel, evaluated against the 2025 reference estimates), the
corrections reduced error less (Gemini 20.2$\to$13.9 linear and 12.0
nested; EXAONE 17.3$\to$14.3 linear and 13.1 nested; the nested selection
within 2024 again chose the quadratic form for six of eight indicators).
Evaluated on the same 2025 cells, the calibrated panel fails to beat simple
references: a target-year grand mean scores 12.0~pp (computed from the entire 2025 age-10-and-over sample, i.e., an oracle-style reference) and a prior-wave (2024) predictor 6.7~pp; the
analogous in-design calibration grand mean scores 11.8~pp on held-out 2024
cells (Table~\ref{tab:realonly}).
This analysis is a backcast of correction coefficients (Sec.~III-F), not a
forecast. Even so, a correction learned on one wave does not transfer
competitively to the next, and the within-wave reduction
(Table~\ref{tab:rq3}, top) reflects contemporaneous error reduction rather
than forward-predictive validity.

\begin{table}[!t]
\caption{RQ3 held-out sex-by-age cell MAE (pp) before and after calibration:
global shift, the post-hoc linear age term, and the nested
selection (form chosen inside the calibration set). The uncorrected column
gives the full-sample cell MAE of Table~\ref{tab:rq2}; on the held-out cells
of the same splits it is 18.3/15.2~pp. Top: 200 stratified within-2024 splits. Middle: temporal holdout (correction learned on 2024,
evaluated on the 2025 panel against the 2025 reference estimates). Bottom:
out-of-time references on the 2025 cells, the target-wave grand mean
(oracle-style) and the prior wave. Both match or beat the temporally calibrated panel,
so the correction reduces raw error across waves without transferring
competitively.}
\label{tab:rq3}
\centering
\setlength{\tabcolsep}{4pt}
\begin{tabular}{lcccc}
\toprule
Model & Uncorrected & Global & Linear age & Nested \\
\midrule
Gemini & 18.3 & 14.2 & 8.3 & 4.9 \\
EXAONE & 15.1 & 8.4 & 6.7 & 4.4 \\
\midrule
Gemini (temporal) & 20.2 & 17.8 & 13.9 & 12.0 \\
EXAONE (temporal) & 17.3 & 14.7 & 14.3 & 13.1 \\
\midrule
Grand-mean (2025) & \multicolumn{4}{c}{12.0} \\
Prior-wave (2024) & \multicolumn{4}{c}{6.7} \\
\bottomrule
\end{tabular}
\end{table}

\begin{table}[!t]
\caption{Calibrated synthetic panel vs.\ real-only estimators on identical
stratified splits (held-out sex-by-age cell MAE, pp; 200 splits per fraction;
$n$ = calibration respondents). Lin.\ = linear age form; Nest.\ = nested
selection; Direct = weighted cell rates from the calibration set (empty cells
fall back to the calibration set's grand mean); Reg.\ = real-only regression (additive in age band and sex); EB =
real-only empirical-Bayes shrinkage of the calibration cell means toward that
regression line (no synthetic data, hence one value per fraction); GM = in-design calibration grand mean; Win = share
of splits in which the nested-calibrated panel beat the better of Direct and
Reg.\ (the share of splits in which each of the four estimators ranks first is reported in Sec.~IV-G). Bottom row: entire-cell holdout (extrapolation to unobserved segments;
direct estimation is undefined there).}
\label{tab:realonly}
\centering
\scriptsize
\setlength{\tabcolsep}{2.2pt}
\begin{tabular}{lccccccccc}
\toprule
& \multicolumn{2}{c}{Lin.} & \multicolumn{2}{c}{Nest.} & \multicolumn{4}{c}{Real-only} & Win \% \\
Fraction & Gem.\ & EXA.\ & Gem.\ & EXA.\ & Direct & Reg.\ & EB & GM & G/E \\
\midrule
1\% ($n{\approx}80$)    & 10.2 & 8.6 & 9.3 & 8.5 & 13.4 & 10.1 & 10.7 & 12.7 & 76/94 \\
2\% ($n{\approx}170$)   & 9.5  & 7.8 & 7.7 & 7.4 & 10.5 & 8.8  & 8.1 & 12.2 & 88/97 \\
3\% ($n{\approx}250$)   & 9.1  & 7.4 & 6.8 & 6.5 & 8.8  & 8.2  & 6.9 & 12.0 & 96/99 \\
5\% ($n{\approx}430$)   & 8.8  & 7.1 & 6.1 & 5.8 & 7.3  & 7.8  & 5.9 & 11.9 & 96/98 \\
10\% ($n{\approx}860$)  & 8.5  & 6.8 & 5.5 & 5.1 & 5.5  & 7.4  & 4.9 & 11.8 & 50/81 \\
20\% ($n{\approx}1{,}730$) & 8.3 & 6.7 & 5.1 & 4.6 & 4.2 & 7.1 & 3.9 & 11.8 & 0/12 \\
30\% ($n{=}2{,}598$)    & 8.3  & 6.7 & 4.9 & 4.4 & 3.6  & 7.1  & 3.5 & 11.8 & 0/0 \\
\midrule
Unobserved cells        & 9.6 & 7.7 & 6.2 & 5.7 & --  & 8.6  & -- & 12.5 & -- \\
\bottomrule
\end{tabular}
\end{table}

\subsection{Comparison to naive baselines}
The uncorrected synthetic panel underperformed a trivial grand-mean baseline at
the cell level (18.3/15.1~pp for Gemini/EXAONE vs.\ 11.6~pp across all eight
indicators): on its own, the persona-imposed segment structure was worse than
assuming no segment variation. Only after calibration (8.3/6.7~pp with the
linear age form; 4.9/4.4~pp with the nested selection) did the
synthetic panel surpass the grand mean (Table~\ref{tab:baseline}). Because the prior-wave baseline covers only the six indicators fielded in 2023
(Sec.~III-G), we compared against it on that common set: the uncorrected panel
scored 13.3/13.7~pp and the grand mean 11.1~pp, versus 3.7~pp for the prior
wave; calibrated (linear form) it reached 6.9/6.8~pp, beating the grand mean but
still short of the prior wave. A prior wave provides no estimates for the two newest indicators (generative
AI, short-form), the one setting in which a calibrated panel could in principle
contribute, subject to the real-only comparison above.

\begin{table}[!t]
\caption{Sex-by-age cell MAE (pp): synthetic vs.\ baselines and ablation (2024,
all eight indicators, identical 14 comparison cells throughout); the uncorrected, demographic-only, and baseline rows are full-sample values on the 14 cells, whereas the calibrated rows are held-out (200-split means). The prior-wave baseline covers only the
six indicators fielded in 2023; on that common set, the values are uncorrected
synthetic 13.3/13.7, grand mean 11.1, prior wave 3.7, and calibrated (linear)
6.9/6.8.}
\label{tab:baseline}
\centering
\begin{tabular}{lcc}
\toprule
Condition & Gemini & EXAONE \\
\midrule
Synthetic (full persona) & 18.3 & 15.1 \\
Synthetic (demographic-only) & 22.3 & 22.1 \\
Grand-mean baseline & \multicolumn{2}{c}{11.6} \\
Prior-wave (2023) baseline (6 common items) & \multicolumn{2}{c}{3.7} \\
Calibrated synthetic (linear age, post-hoc) & 8.3 & 6.7 \\
Calibrated synthetic (nested selection) & 4.9 & 4.4 \\
\bottomrule
\end{tabular}
\end{table}

\subsection{Persona conditioning ablation}
Replacing the persona narrative with demographic-only conditioning degraded
performance for both models: the sex-by-age cell MAE rose from 18.3 to 22.3~pp
(Gemini) and from 15.1 to 22.1~pp (EXAONE; Table~\ref{tab:baseline}), and the
overall RQ1 MAE rose from 17.0 to 18.3~pp and from 13.8 to 21.3~pp. The
narrative therefore adds useful signal beyond demographics, even though it does
not by itself surpass the naive baseline. This mirrors evidence that
demographic-only agents underperform agents grounded in richer individual
self-reports \cite{parkjs}. It contrasts with a Seoul Survey benchmark in which narrative conditioning worsened fit relative to demographics alone~\cite{parkseoul}; the persona source, models, and item domain all differ, so the two results are not directly comparable.

\subsection{Response stability}
A variance decomposition over $K{=}5$ responses per persona yielded an ICC of
0.84 (Gemini) and 0.34 (EXAONE): generation noise accounted for about 16\% of
the variance for Gemini versus 66\% for EXAONE, consistent with EXAONE's greater
temperature sensitivity. Aggregate estimates were stable nonetheless:
across three independent full passes of the EXAONE panel, each evaluated on
its own valid personas, the overall MAE spanned only 13.7--13.8~pp (sex-by-age cell MAE 15.0--15.2), and averaging each
persona's three responses yielded an overall MAE of 13.7~pp,
indistinguishable from single-pass results. A hierarchical bootstrap that resamples households, then personas, then
one of each persona's three EXAONE responses (8{,}158 personas valid in all
three passes) gave a 95\% CI for the overall MAE of [13.2, 14.3]~pp (width
1.17~pp), against [13.2, 14.4] (width 1.18) for the single-pass design-based
bootstrap of Sec.~III-E on the same age-10-and-over sample; averaging the three
responses first narrowed it to [13.2, 14.3] (width 1.05). The same check for
Gemini, on the subsample with three fixed-order responses
(1{,}144 sampled, 1{,}141 valid in all arms; Sec.~IV-K), gave [16.1, 17.6]~pp
(width 1.45) against [16.0, 17.4] (width 1.44) for the single-pass bootstrap and
[16.2, 17.5] (width 1.30) with the responses averaged. In both models the
hierarchical interval is at most 0.01~pp wider than the single-pass one, so at
this resolution generation stochasticity did not widen the aggregate interval;
this is an observation for these runs and settings, not a general bound on
generation uncertainty. The
wider Gemini interval reflects the subsample's size (1{,}141 against 8{,}158
personas); those
fixed-order responses were also generated in September 2026 rather than July,
so they carry any serving drift between the two dates. Because generation noise averaged out over the 200--600 personas per cell (fewer for the two OTT-gated items, Sec.~III-D),
one response per persona was sufficient for stable aggregate estimates in these runs.

\subsection{Questionnaire order}
To test the order component of the questionnaire-context effects noted in
Sec.~III-D, the 1{,}144-persona stratified subsample of the wording
experiment answered the 2024 questionnaire again with Gemini on 3 September 2026: three
times in the fixed codebook order (F1--F3, generated on the same day, which serve as
generation-noise references and as the repeated responses used in
Sec.~IV-J) and once with an independent random permutation of the 36 items
for each persona (R1); model, temperature, prompts, wording, and options were
unchanged, and 1{,}141 personas were valid in all four arms.
Table~\ref{tab:order} compares post-stratified rates. Randomizing the order
shifted five of the eight indicators significantly
(Holm-adjusted bootstrap $p\le0.012$): short-form use rose from 33.4\% to
60.1\% ($+26.7$~pp, 95\% CI [21.6, 31.0]), subscription from 40.1\% to
49.8\% ($+9.6$), generative-AI use from 41.1\% to 46.9\% ($+5.8$), SNS use
from 49.5\% to 53.2\% ($+3.7$), and metaverse use from 1.6\% to 3.3\%
($+1.6$). OTT and messenger use showed no significant change, and among the OTT-gated
personas YouTube use was 100\% in every arm. Four of the five
shifts (short-form, subscription, generative-AI, and SNS use) also exceed the
largest difference between any two fixed-order passes (2.8~pp); the metaverse
shift is significant but within that range, so it is not
separable from pass-to-pass variation by magnitude alone. The
share of personas whose answer changed was up to 14\% between fixed-order passes
versus up to 32\% under randomization. The 5-point construct means rose by
0.05--0.30 under random order, so interspersing the Likert battery among the
yes/no items weakened Gemini's low anchoring somewhat. Item order is
therefore a real design factor for a single-call questionnaire, with effects
that reach the size of the headline error for individual items. It does not,
however, change the paper's conclusions: on this subsample the overall MAE
against the 2024 reference moved from 16.8 (fixed) to 14.8~pp (random),
every bias direction was preserved (generative-AI and subscription use
overestimated, short-form and OTT use underestimated), and the short-form
gap remained 9.7~pp wide under randomized order (60.1 against the survey's 69.8\%). Both wording and order therefore matter. The two are not on a common scale, because only the order arms carry the skip logic, so we do not rank them; their interaction was not tested (no neutral-wording, random-order arm), and part of the fixed-order collapse may reflect the item's placement immediately after the OTT and YouTube items. We report
the main panel under the fixed codebook order and treat this experiment as
bounding the order component of the context effects listed in Sec.~VI.

\begin{table}[!t]
\caption{Randomized-order regeneration (Gemini; 1{,}141 personas valid in all
arms; post-stratified rates, \%). F1/F2 = fixed codebook order, first and
second pass (a third pass, F3, lies within 1.6~pp of both on every indicator; package sheet); R1 = random order per persona. The change column is R1 minus F1, computed from unrounded rates, with a
persona-level bootstrap 95\% CI and Holm-adjusted $p$ (eight indicators).
Changed = share of personas whose answer differed between F1 and R1 / between
F1 and F2. Survey = 2024 weighted rate (ages 10 and over; YouTube and short-form among OTT users).}
\label{tab:order}
\centering
\scriptsize
\setlength{\tabcolsep}{1.3pt}
\begin{tabular}{lcccclcc}
\toprule
Indicator & Survey & F1 & F2 & R1 & $\Delta$ [95\% CI] & Holm $p$ & Changed \\
\midrule
AI use & 13.8 & 41.1 & 40.5 & 46.9 & $+5.8$ [3.9, 7.8] & 0.008 & 11.1/5.3 \\
OTT & 89.2 & 64.1 & 63.5 & 62.7 & $-1.5$ [$-3.4$, 0.5] & 0.33 & 10.3/6.2 \\
YouTube & 93.2 & 100.0 & 100.0 & 100.0 & 0.0 [0.0, 0.0] & 1.00 & 0.0/0.0 \\
Short-form & 69.8 & 33.4 & 31.9 & 60.1 & $+26.7$ [21.6, 31.0] & 0.008 & 32.1/13.6 \\
SNS & 61.4 & 49.5 & 46.7 & 53.2 & $+3.7$ [1.6, 5.8] & 0.012 & 10.5/7.1 \\
Messenger & 92.5 & 98.2 & 98.2 & 97.5 & $-0.7$ [$-1.5$, 0.1] & 0.23 & 2.0/1.3 \\
Metaverse & 5.5 & 1.6 & 1.9 & 3.3 & $+1.6$ [0.8, 2.4] & 0.010 & 3.0/2.0 \\
Subscription & 22.9 & 40.1 & 39.0 & 49.8 & $+9.6$ [7.5, 11.7] & 0.008 & 14.7/11.3 \\
\midrule
Overall MAE (pp) & -- & 16.8 & 17.1 & 14.8 & & & \\
\bottomrule
\end{tabular}
\end{table}

\section{Discussion}
\textbf{Limited but structured validity.} The synthetic persona panel reproduces
the target distributions only imperfectly (RQ1 MAE 14--19~pp), and the error
is of the same magnitude at the subgroup level (five-axis segment MAE
14--18~pp). It is
not random noise: it decomposes into identifiable causes, which makes it
diagnosable and, within a wave, partially correctable.

\textbf{Temporal misalignment: consistent but partial.} The generative-AI estimates sit
near more recent adoption levels, and aligning the reference year and item wave
removes most of the AI overestimation, a pattern consistent with temporal
misalignment between the generation-time knowledge (model and persona source)
and the survey year, though its exact source cannot be isolated
(Sec.~IV-E). The mismatch does not, however,
explain errors in the opposite direction: short-form use is underestimated, and the gap
\emph{widens} as real adoption rises. The overall MAE therefore does not improve under
alignment: the temporal-misalignment pattern is confined to specific
indicators and does not account for the broad misfit.

\textbf{Model-specific bias signatures.} Gemini's errors amplify a
``digital = young'' pattern and anchor responses low on the OTT, short-form,
and SNS items, driving those rates down through the 40s to 60s bands before the gap narrows
again at 70 and over, while messenger use moves the other way ($+23$~pp at 70
and over). EXAONE's binary
errors are per-indicator offsets whose direction varies by item; its upward
tendency is carried by the Likert constructs (3.03 vs.\ 2.83) and the
yes-selection rate (59.5 vs.\ 55.9\%), so we call it acquiescence-consistent
(no reverse-keyed items were fielded, so it is not confirmed acquiescence). These distinct
signatures echo prior findings on response homogenization~\cite{park} and
acquiescence~\cite{braun} and imply that no single mitigation recipe will serve
all models. They also speak to the design question posed in Sec.~III-C: the
Korea-developed model did track Korean response levels more closely (MAE 13.8
vs.\ 17.0~pp) but showed lower response stability (ICC 0.34 vs.\ 0.84) and
weaker construct ordering ($r=0.500$ vs.\ 0.788). Because the two deployed
systems differ in scale, training data, and serving stack, this is a
descriptive contrast between systems, not an isolated effect of locale
specialization.

\textbf{Framing sensitivity.} The short-form item, framed as an ``OTT service,''
caused synthetic endorsement to collapse despite near-ceiling YouTube estimates,
showing that survey wording can dominate synthetic validity and must therefore be
standardized and reported. We could not field a human split-ballot arm: the
study is a secondary-data design under an institutional review board (IRB)
exemption that covers no new data collection with human participants, so the wording experiment establishes LLM framing \emph{sensitivity} rather than whether human respondents would move by a comparable amount. Published human split-ballot experiments on comparable items are the natural benchmark for a follow-up. The within-survey consistency check (short-form use conditional on YouTube use, which holds at
73\% among real respondents but 42.5\% for Gemini and 58.6\% for EXAONE in the 2024 run) shows
that the models' conditional rates sit far below the survey's under the original wording; whether their reaction to the wording exceeds a human respondent's cannot be judged from this comparison.
The randomized-order regeneration (Sec.~IV-K) adds a second boundary: randomizing the order alone recovered 26.7~pp and left a 9.7~pp gap under the codebook wording. Only the order arms carry the skip logic, so the two effects are not on a common scale and are not ranked; their interaction was not tested.

\textbf{Reconciling with the U.S.\ evidence.} Our negative result need not
contradict the positive silicon-sampling findings for U.S.\
opinions~\cite{argyle}. Political attitudes are well represented in
training corpora and change slowly, whereas fast-moving technology-adoption
behavior in a non-English market combines thinner cultural representation with
rapid drift, the conditions under which stereotype priors and framing
literalism dominate. This contrast is consistent with synthetic-panel validity
being domain- and locale-contingent rather than a general property of LLMs,
although a single-country, single-domain study cannot establish that.

\textbf{Added value is conditional, and narrower than calibration alone
suggests.} Three baseline comparisons bound the panel's value.
First, the uncorrected panel performs \emph{worse} than a no-segment-information grand-mean baseline at the segment level, because the persona-conditioned variation is wrong in a
model-specific way: stereotyped age structure for Gemini, per-indicator level
offsets for EXAONE. Second, although calibration corrects that variation
and beats the grand mean, the calibration sample itself is real data: used
directly, it estimates the same cells more accurately (3.6 vs.\ 4.9/4.4~pp
with the nested correction, 8.3/6.7~pp with the linear form, at the 30\%
fraction; Sec.~IV-G). The calibrated panel keeps a margin only when the real
calibration sample is smaller than the crossover sizes of Sec.~IV-G, and even
there the margin over a real-only shrinkage estimator is at most 2.3~pp. The
unobserved-segment advantage holds for both models under the nested
correction, but only for EXAONE once the teen cells are excluded. Third, under
a one-year gap the correction does not transfer competitively: the temporally calibrated
panel (12.0--14.3~pp) fails to beat a same-year grand-mean reference
(12.0~pp), although that reference is computed from the entire target-year
sample and so is oracle-style (Sec.~IV-G). The within-wave gain is contemporaneous error reduction
in a narrow niche, not forward-predictive validity. Response stability is also
model-dependent (Sec.~IV-J): aggregate estimates are unaffected, but
persona-level applications of noisier models require multiple responses.

\textbf{Practical implication.} Even when aggregate means are close, specific
segments (e.g., the middle and older age bands on OTT and SNS use) can be misestimated by a wide margin, so direct subgroup use
is risky. Unlike statistical real--synthetic combination~\cite{wang} or
distribution-matched fine-tuning~\cite{cao}, our correction is simple and
training-free. Its contribution is the selection of the correction form
\emph{inside} the calibration set rather than by inspection, a procedure that
chose an age term for both models, together with a characterization of its limits
(Sec.~IV-G). Any operational use should be benchmarked against real-only
estimators before deployment. A standalone forecasting use, the pre-diffusion
window~\cite{rogers} in technology-forecasting terms, remains a hypothesis our
data do not establish; the out-of-time analysis suggests that such use would
also have to overcome non-transferable calibration.

\textbf{Where the niche lies in practice.} The data-scarce regime in which a
calibrated panel retains an edge (fewer than a few hundred real responses in
total, at most about 30 per sex-by-age cell, and even there by at most about 2~pp over
a well-chosen real-only estimator) can arise in several settings: (a) the
early diffusion of an emerging technology, when no survey wave has yet fielded
the item and only a small pilot exists; (b) sensitive or hard-to-reach topics,
where probability samples are necessarily small; and (c) budget-constrained
studies that can afford a few hundred interviews but not a full wave. In each
case the calibrated panel is a stopgap that must be re-benchmarked as soon as
more real data arrive, because its margin disappears at the crossover sizes of
Sec.~IV-G.

\textbf{A practical workflow.} The findings translate into a decision rule
(Fig.~\ref{fig:flow}). (0)~First ask whether a real probability sample for the
target wave exists or can be collected. Direct estimation matches or beats the calibrated panel from
roughly 860 respondents for Gemini (a tie at that size) and 1{,}730 for EXAONE, the first tested sizes at which
the calibrated panel no longer won; its partially pooled version, which shrinks cell means
toward the real-only regression line, wins from a few hundred responses
earlier still (Sec.~IV-G). Above those sizes the synthetic-target
shrinkage variant improves on its real-only counterpart by at most 0.3~pp for
Gemini and 0.7~pp for EXAONE, and the synthetic panel should not be used for estimation.
Between the two thresholds, 250 to 860 respondents for Gemini and 430 to
1{,}730 for EXAONE, direct estimation has not yet won but a partially pooled
real-only estimator already leads the calibrated panel in this study, by
0.2--0.7~pp at the tested sizes (Table~\ref{tab:realonly}); a user in that range
should benchmark against that estimator and expect it to win narrowly. The synthetic panel enters in one of two ways. For an observed segment it enters by default only when the real sample is
scarce, below roughly 250 responses for Gemini and 430 for EXAONE. For an entirely unobserved segment, where direct
estimation is undefined, it enters only after a held-out-cell evaluation shows
it beating a real-only regression; that evaluation passed for both models on
the 14 cells but reversed for Gemini once the teen cells were removed (6.7
against 6.2~pp; Sec.~IV-G), and where it fails the fallback is a real-only
regression fitted on the observed cells, not the calibrated panel. Five steps then apply.
(1)~Generate the panel with locale-appropriate persona narratives
rather than demographics alone (Sec.~IV-I).
(2)~Use whatever real responses exist to diagnose the model's bias signature
(age-slope, response-style, and framing checks as in Secs.~IV-C and IV-D).
(3)~Apply a signature-matched correction whose form is selected inside the
calibration data (Sec.~III-F).
(4)~Benchmark the calibrated panel against direct estimation, a real-only
regression, and grand-mean and prior-wave references, and discard it wherever
it fails to beat them; do not reuse a correction across waves without
re-benchmarking.
(5)~Before any persona-level application, verify response stability (ICC) and
switch to multi-response generation if the model is noisy.

\Figure[t!](topskip=0pt, botskip=0pt, midskip=0pt)[width=0.8\textwidth]{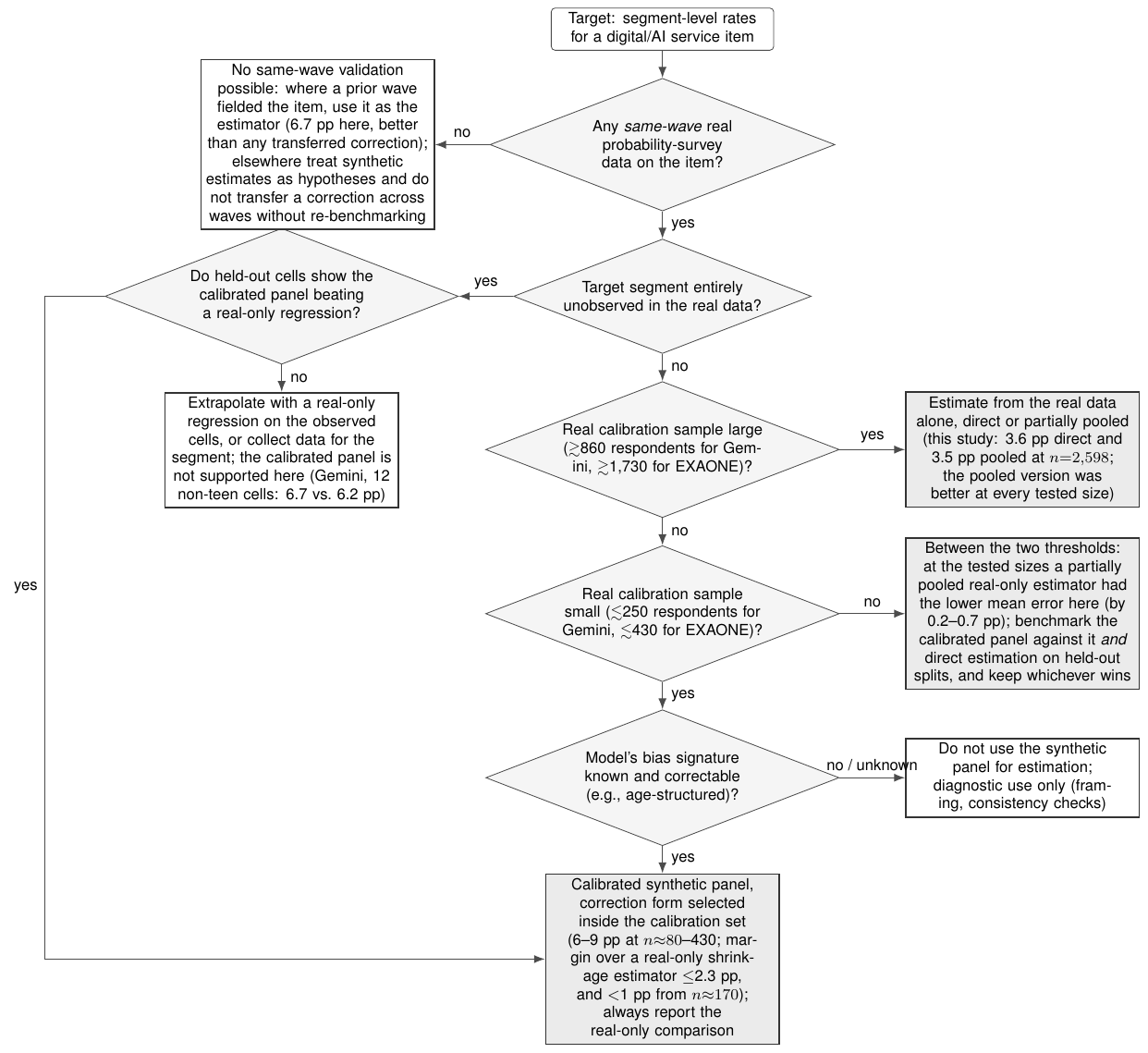}{Decision flow for using a calibrated synthetic panel against real-only estimation, with the error levels observed in this study (held-out sex-by-age cell MAE, pp). The upper thresholds are the first tested sizes at which the calibrated panel no longer beat direct estimation (a tie for Gemini at 860), and the lower ones the largest tested sizes at which the calibrated panel still beat a real-only shrinkage estimator (Sec.~IV-G). On unobserved cells the calibrated panel beat the real-only regression for both models on the 14 cells (5.7--6.2 vs.\ 8.6~pp) but only for EXAONE on the 12 non-teen cells (4.8 vs.\ 6.1~pp; Gemini 6.7 vs.\ 6.2~pp).\label{fig:flow}}

\section{Conclusion and Limitations}
In answer to the research questions, (RQ1) overall agreement is limited, with
an MAE of 14--19~pp and systematic, model-specific bias rather than noise. (RQ2) Error
concentrates in particular indicator-by-segment combinations, deepening through
the middle and older bands on OTT, short-form, and SNS use under Gemini, with a between-group signed-error range ($R_e$) of nearly 50~pp for Gemini
(49.6; EXAONE 34.7; 39.5/31.2 without the non-comparable teen cells). (RQ3) Calibration helps only in a narrow sense. A signature-matched correction whose form is
selected inside the calibration set cuts the sex-by-age cell error by about 70\%
(18.3/15.1 to 4.9/4.4~pp), but a real-only estimator is more
accurate once a few hundred real responses exist (direct estimation beyond roughly 430--1{,}730, a real-only shrinkage estimator beyond roughly 250--860,
with margins of at most about 2~pp below that), a synthetic-informed shrinkage estimator improves on its real-only counterpart by at most 0.7~pp, and
the correction reduces raw error across waves but does not transfer competitively (out-of-time, in a backcast from 2024 to 2025, it trails both a
grand-mean and a prior-wave reference). Within this setting, synthetic persona panels are not
substitutes for real measurement: their defensible uses are diagnostic, and
operationally they are confined to small calibration samples and entirely
unobserved segments, where the advantage held for both models on the 14
sex-by-age cells but for EXAONE alone without the teen cells. The findings are scoped to Korea, the digital and AI service-use
domain, two contrasting LLMs, and the 2024--2025 window; generalization beyond
these bounds (to other cultures, domains, models, or periods) remains an open
question. Several limitations apply. (1) The metrics are restricted to binary use
and 5-point constructs, so open-ended responses and behavior logs are not
covered. (2) The main analysis uses one response per persona.
Generation variance was quantified on a subsample (ICC 0.84 for Gemini, 0.34
for EXAONE), and aggregate robustness was verified by three independent full
passes of the noisier model (headline MAE varying by about 0.1~pp) and by a
hierarchical bootstrap (Sec.~IV-J); persona-level analyses would still require
multiple responses. (3) The 2025 item set lacks constructs, restricting the
temporal-alignment analysis to binary indicators. (4) The education and
employment segment axes rely on heuristic crosswalks between persona
attributes and KISDI categories (the region mapping was validated empirically
against population shares, but the education and employment mappings are
approximate). Misclassification can attenuate or exaggerate the measured
segment error along those axes, and without a misclassification model the
direction and magnitude of that distortion are unknown, so the education and employment rows of
Table~\ref{tab:rq2} should be interpreted cautiously. (5) The 2025 item set contains 12 items
against 36 in 2024, so the \emph{additional} widening of the short-form gap
under matched-wave regeneration is partly confounded with questionnaire length
and context. The core framing effect is already present in the length-matched
2024 run and is confirmed by the wording-controlled regeneration
(Sec.~IV-D), so this confound does not affect the framing conclusion. That
experiment had no human split-ballot arm, because the secondary-data design
and its IRB exemption cover no new human data collection, so it establishes
that the models are framing-sensitive, not that human respondents would move by a comparable amount. Each arm also used a single generation call, so the EXAONE arms dropped
12.8--15.3\% of personas and that contrast is directional evidence only (Sec.~IV-D). The experiment's per-persona pairing was not retained, so its intervals are
independent-sample rather than paired, and they are conservative only if the
within-persona correlation is nonnegative. (6) The
synthetic teen cell contains only age-19 personas, whereas the survey's teen
band spans ages 10--19; age-segment inference is therefore qualified by the
teen-excluded results, and a like-for-like comparison with the survey's
19-year-olds shows that roughly half (Gemini) or a quarter (EXAONE) of the apparent teen error is a band
artifact (Sec.~IV-B). (7) No reverse-keyed items were fielded, so EXAONE's
upward shift is labeled acquiescence-consistent rather than confirmed
acquiescence. (8) The paper reports many indicator-by-axis comparisons
without a global multiplicity correction; the analysis-status table
(Sec.~III-H) and the family-level Holm-adjusted results for the principal model comparison, the wording experiment, and the randomized-order experiment delimit what is confirmatory, and
RQ3 in its entirety is post-hoc and not preregistered. (9) The temporal analyses are
retrospective (the 2024-to-2025 comparison is a reference-year swap, and the
calibration transfer is a backcast of coefficients onto responses generated in
2026), so nothing here demonstrates pre-diffusion forecasting performance,
and the three temporal channels (model knowledge, persona-source timing,
diffusion reasoning) remain confounded (Sec.~IV-E). (10) Item order was fixed
and all items shared one context window; the randomized-order regeneration
(Sec.~IV-K) bounds the order component (up to 27~pp for a single item but
2.0~pp on the overall MAE, with all bias directions preserved) but not other
context effects such as questionnaire length or within-call consistency
pressure, and it was run for Gemini only. (11) Construct-level
claims are restricted to agreement in means, and the correlation-structure
diagnostics (Table~\ref{tab:concorr}) show that psychometric equivalence does
not hold~\cite{lukauskas}. (12) The design-based intervals reproduce weighting and
household clustering but not the survey's 56-stratum design or its
first-stage clustering by enumeration district (the microdata carry no
district identifiers), so their survey-sampling component is approximate. (13) The survey reference is a long-running panel, so panel conditioning may remain in the reported responses as a residual measurement error that weighting does not remove. (14) The EXAONE 2025 exclusions were concentrated in one cell (men in their 40s, 7.5\%), whose 2025 estimates therefore carry a worst-case exclusion bound of 7.5~pp on the six full-denominator items and 8.6~pp on the two OTT-gated items. Under the OTT gate of Sec.~III-D, the 2025 Gemini cell for women aged 70 and over retains only 7 personas (22 for men), so its YouTube and short-form estimates rest on a handful of responses and are unstable. (15) Four scope limits follow from the design. The calibration analysis covers
the eight binary indicators only, not the constructs. Segment error on the
education, employment, and region axes is post-stratified on sex-by-age only,
so persona-pool composition within those groups contributes to those rows
beyond the crosswalk issue in~(4). The set of five candidate correction forms
was itself fixed after inspecting the 2024 data, and the nested selection
removes the choice among them but not the choice of the set; at the smallest
fractions that selection fits up to four coefficients, the intercept included,
on about 40 respondents.
The closed-weight model is served by its provider and may drift, so its
responses, archived in full, cannot be regenerated exactly. Future work should examine framing standardization against published human
split-ballot experiments and prospective validation on a wave fielded after
generation, together with stereotype-mitigation prompting, response-style
correction, and multi-response uncertainty quantification.

\appendix
\section*{Appendix: Supplementary Tables and Prompt Structure}
\begin{table}[!t]
\caption{Teen-cell-excluded sensitivity: full sample / teen cells excluded
(12 cells). Orderings are preserved except the Gemini entire-cell holdout (6.2 vs.\ 8.6~pp on 14 cells; 6.7 vs.\ 6.2~pp on the 12 non-teen cells); $R_e$ shrinks while headline error levels do not.}
\label{tab:teenexcl}
\centering
\scriptsize
\setlength{\tabcolsep}{3pt}
\begin{tabular}{lcc}
\toprule
Metric (pp) & Gemini & EXAONE \\
\midrule
RQ1 MAE, 2024 & 17.0 / 17.7 & 13.8 / 13.7 \\
RQ1 MAE, 2025 & 19.4 / 20.4 & 15.2 / 15.2 \\
RQ2 sex-by-age cell MAE & 18.3 / 17.6 & 15.1 / 15.0 \\
$R_e$ & 49.6 / 39.5 & 34.7 / 31.2 \\
Five-axis MAE range & 15.2--18.3 / 15.6--17.5 & 13.9--15.6 / 13.6--16.1 \\
Calibrated (linear age, post-hoc) & 8.3 / 7.6 & 6.7 / 5.7 \\
Calibrated (nested selection) & 4.9 / 4.8 & 4.4 / 3.7 \\
Real-only direct (30\%) & \multicolumn{2}{c}{3.6 / 3.2} \\
Unobserved cells, nested & 6.2 / 6.7 & 5.7 / 4.8 \\
Unobserved cells, real-only reg. & \multicolumn{2}{c}{8.6 / 6.1--6.2} \\
Out-of-time calibrated (linear) & 13.9 / 13.6 & 14.3 / 13.6 \\
Grand-mean baseline & \multicolumn{2}{c}{11.6 / 12.5} \\
Demographic-only ablation & 22.3 / 20.3 & 22.1 / 21.7 \\
\bottomrule
\end{tabular}
\end{table}
\begin{table}[!t]
\caption{Gemini: signed error by age band (pp; synthetic $-$ survey, 2024; YouTube and short-form among OTT users).}
\label{tab:ageerr-g}
\centering
\footnotesize
\begin{tabular}{lrrrrrrr}
\toprule
Indicator & 10s & 20s & 30s & 40s & 50s & 60s & 70s+ \\
\midrule
AI use & $+71$ & $+47$ & $+37$ & $+29$ & $+13$ & $+4$ & $+2$ \\
OTT & $+2$ & $+1$ & $-3$ & $-11$ & $-40$ & $-64$ & $-40$ \\
YouTube & $+9$ & $+8$ & $+6$ & $+9$ & $+6$ & $+5$ & $+3$ \\
Short-form & $+11$ & $-15$ & $-51$ & $-61$ & $-60$ & $-47$ & $-20$ \\
SNS & $+34$ & $+6$ & $-9$ & $-31$ & $-35$ & $-26$ & $-4$ \\
Messenger & $+5$ & $+1$ & $+1$ & $+1$ & $+1$ & $+6$ & $+23$ \\
Metaverse & $-8$ & $-9$ & $-8$ & $-2$ & $-1$ & 0 & 0 \\
Subscription & $+42$ & $+35$ & $+35$ & $+20$ & 0 & $-4$ & $-1$ \\
\bottomrule
\end{tabular}
\end{table}
\begin{table}[!t]
\caption{EXAONE: signed error by age band (pp; synthetic $-$ survey, 2024; YouTube and short-form among OTT users).}
\label{tab:ageerr-e}
\centering
\footnotesize
\begin{tabular}{lrrrrrrr}
\toprule
Indicator & 10s & 20s & 30s & 40s & 50s & 60s & 70s+ \\
\midrule
AI use & $+45$ & $+28$ & $+27$ & $+28$ & $+19$ & $+13$ & $+10$ \\
OTT & $-4$ & $-4$ & $-8$ & $-11$ & $-12$ & $-15$ & $+21$ \\
YouTube & $+8$ & $+7$ & $+5$ & $+7$ & $+2$ & $-2$ & $-7$ \\
Short-form & $-9$ & $-16$ & $-23$ & $-19$ & $-13$ & $-7$ & $+14$ \\
SNS & $+29$ & $+3$ & $-3$ & $-2$ & $+8$ & $+21$ & $+42$ \\
Messenger & $-8$ & $-10$ & $-17$ & $-24$ & $-33$ & $-41$ & $-16$ \\
Metaverse & $-5$ & $+9$ & $+9$ & $+12$ & $+11$ & $+10$ & $+6$ \\
Subscription & $+20$ & $+1$ & $+16$ & $+25$ & $+27$ & $+26$ & $+26$ \\
\bottomrule
\end{tabular}
\end{table}
\begin{table*}[!t]
\caption{Linear age correction (Sec.~III-F): fitted intercept $\beta_0$ and
slope $\beta_1$ (pp; pp per decade band) per indicator, mean and 2.5/97.5
percentiles over the 200 calibration splits; residual SD of the full-data fit
over the 14 cells; and the split-to-split half-width of the corrected cell
estimate (HW, pp, averaged over cells).}
\label{tab:coef}
\centering
\footnotesize
\setlength{\tabcolsep}{5pt}
\begin{tabular}{llcccc}
\toprule
Model & Indicator & $\beta_0$ [2.5, 97.5] & $\beta_1$ [2.5, 97.5] & Resid.\ SD & HW \\
\midrule
Gemini & AI use & 62.9 [58.7, 67.0] & $-11.3$ [$-12.2$, $-10.4$] & 5.9 & 2.3 \\
Gemini & OTT & 9.6 [7.7, 12.1] & $-10.6$ [$-11.4$, $-9.9$] & 12.1 & 1.7 \\
Gemini & YouTube & 9.4 [5.9, 13.0] & $-0.9$ [$-1.7$, $-0.2$] & 3.1 & 2.1 \\
Gemini & Short-form & $-16.8$ [$-21.5$, $-11.3$] & $-6.0$ [$-7.4$, $-4.7$] & 23.1 & 3.5 \\
Gemini & SNS & 12.4 [6.9, 18.1] & $-7.3$ [$-8.7$, $-6.0$] & 19.7 & 3.3 \\
Gemini & Messenger & $-1.3$ [$-3.5$, 1.3] & 2.2 [1.4, 2.9] & 5.7 & 1.9 \\
Gemini & Metaverse & $-9.2$ [$-13.9$, $-4.8$] & 1.7 [0.7, 2.7] & 1.7 & 2.2 \\
Gemini & Subscription & 44.3 [39.3, 49.1] & $-8.7$ [$-9.8$, $-7.6$] & 6.7 & 2.8 \\
\midrule
EXAONE & AI use & 40.1 [35.9, 44.2] & $-5.2$ [$-6.1$, $-4.3$] & 5.4 & 2.3 \\
EXAONE & OTT & $-10.2$ [$-12.0$, $-7.6$] & 1.8 [0.9, 2.4] & 11.0 & 1.7 \\
EXAONE & YouTube & 10.2 [6.7, 13.8] & $-2.4$ [$-3.2$, $-1.7$] & 3.8 & 2.1 \\
EXAONE & Short-form & $-20.8$ [$-25.5$, $-15.2$] & 3.5 [2.1, 4.8] & 9.6 & 3.5 \\
EXAONE & SNS & 4.6 [$-0.9$, 10.4] & 3.1 [1.7, 4.4] & 15.2 & 3.3 \\
EXAONE & Messenger & $-10.4$ [$-12.6$, $-7.9$] & $-3.7$ [$-4.5$, $-3.0$] & 9.0 & 1.9 \\
EXAONE & Metaverse & 3.6 [$-1.1$, 8.0] & 1.3 [0.3, 2.3] & 5.0 & 2.2 \\
EXAONE & Subscription & 11.9 [7.0, 16.8] & 2.8 [1.7, 3.9] & 6.8 & 2.8 \\
\bottomrule
\end{tabular}
\end{table*}
\begin{table*}[!t]
\caption{Construct correlation structure (eight 5-point constructs, 2024):
mean and maximum absolute difference between the synthetic and survey
inter-construct correlations (28 pairs; survey correlations weighted),
correlation between the two correlation vectors, mean absolute correlation,
ratio of construct SDs (synthetic/survey), and MAE of construct means (1--5 scale). The construct MAE in this table compares unweighted synthetic construct means with the weighted survey means on the age-10-and-over sample (0.37/0.39); the value in Sec.~IV-A (0.373/0.380) post-stratifies the synthetic means and uses the full survey sample, so the two are not the same statistic.}
\label{tab:concorr}
\centering
\footnotesize
\setlength{\tabcolsep}{6pt}
\begin{tabular}{lcccccc}
\toprule
Model & Mean $|\Delta r|$ & Max $|\Delta r|$ & $r$(vectors) & Mean $|r|$ syn./survey & SD ratio (range) & Mean MAE \\
\midrule
Gemini & 0.085 & 0.252 & 0.01 & 0.70 / 0.68 & 0.71 (0.48--0.98) & 0.37 \\
EXAONE & 0.108 & 0.320 & 0.35 & 0.62 / 0.68 & 0.85 (0.65--0.96) & 0.39 \\
\bottomrule
\end{tabular}
\end{table*}
Table~\ref{tab:teenexcl} reports the teen-cell-excluded sensitivity analysis
referenced in Secs.~IV-B and VI: every headline metric recomputed on the 12 cells that exclude the two synthetic teen cells, the least population-comparable cells.
Tables~\ref{tab:ageerr-g} and~\ref{tab:ageerr-e} report the cell-level signed
error (synthetic $-$ survey, pp, 2024, sexes pooled: the survey band rate is weighted, and the synthetic band rate pools the two sex cells without weights) underlying the RQ2 and
bias-signature analyses.

\textbf{Prompt structure.} Every request carried a system instruction and a user prompt, both in Korean (verbatim templates in the package's \texttt{PROTOCOL.md}; English rendering here). System instruction: ``You are an individual in Korea with the following characteristics: [persona block]. Answer the survey consistently from this person's standpoint and honestly, like a real person. Answer each item with exactly one of the listed option numbers.'' The persona block concatenates five labeled fields: [Basic information] sex, age, residence, occupation, and education in the source record's own wording; [Summary]; [Background]; [Hobbies and interests]; and [Culture and media orientation]. The user prompt lists every item as ``[item code] item text'' (original KISDI wording) with its options as number--label pairs and ends with the instruction to output one JSON object mapping item codes to option numbers, with an example; EXAONE requests add ``Output the JSON object only.'' The demographic-only ablation replaces the persona block with ``sex [male/female], age band [teens to 70s and over].'' Live-call (EXAONE) retries append the previous format error to the system instruction; batch (Gemini) retries resubmit unchanged.

The age gradients discussed in Sec.~IV-C are visible in
Tables~\ref{tab:ageerr-g} and~\ref{tab:ageerr-e} (the Sec.~IV-C slopes are 2025 matched-wave fits; the 2024 slopes are in Table~\ref{tab:coef}): Gemini's generative-AI overestimation shrinks from $+71$~pp
among teens to $+2$~pp among those in their 70s, while its OTT, short-form,
and SNS errors deepen into the middle and older bands and ease in the
70s-and-over band; EXAONE's errors are flatter across age but shifted in
level.

\section*{Ethics Statement}
This study is a secondary analysis of de-identified microdata from the KISDI Korea
Media Panel Survey and of synthetically generated persona responses; no new human
subjects were recruited. The study was granted an exemption from review by the
Institutional Review Board of Sungkyunkwan University (exemption determination no.\ SKKU
2026-07-021). No personally identifiable information is used or released.

\section*{Data and Code Availability}
The analysis and generation code, prompt protocol (system instructions, item
text, JSON schema, and skip logic), generation run logs and the retained batch request payloads
(Sec.~III-D), persona-selection seeds and
dataset version, raw synthetic responses with per-group failure rates, category
crosswalks, weighting and calibration code, and a registration correspondence table (\texttt{PREREGISTRATION\_MAP.md}) are archived in a public reproducibility package at
\url{https://github.com/howardkim1977/persona-validation-repro} (MIT license,
with SHA-256 manifests), with immutable DOI-stamped snapshots at
\url{https://doi.org/10.5281/zenodo.21397425} (release v1.0) and \url{https://doi.org/10.5281/zenodo.22324669} (release
v1.23). A quick-start
script (\texttt{quickstart.py}) verifies the package manifest, prints the
headline table values from the archived aggregate workbook, and regenerates the panels of Figs.~1--4 (Fig.~5 is a TikZ source compiled separately) without KISDI access. The KISDI microdata are
access-restricted and available from KISDI upon application; the
Nemotron-Personas-Korea dataset is publicly available on Hugging
Face~\cite{nemotron}. Individual-level KISDI-derived data and API keys are
excluded from the package.
\section*{Acknowledgment}
During the preparation of this article, the authors used Claude (Anthropic) to
improve the grammar, clarity, and readability of text written by the authors
and to assist in developing, refactoring, and debugging portions of the
analysis and figure-generation code provided in the reproducibility package.
No generative AI system was used to formulate the research questions, design
the study, generate or select the reference data, choose the analytical
methods, interpret the results, or derive the scientific claims. The authors
reviewed and edited all AI-assisted text, inspected and tested all AI-assisted
code, verified the reported numerical results against the underlying data, and
take full responsibility for the content of the article and the accompanying
code. No AI-generated text or images appear in the article itself; the AI-assisted code is confined to portions of the analysis and figure-generation code in the accompanying reproducibility package, and the authors verified the figures that those scripts produce.

\begin{IEEEbiography}[{\includegraphics[width=1in,height=1.25in,clip,keepaspectratio]{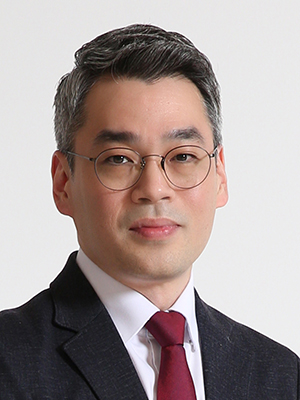}}]{Howard Kim}
received the B.A. and M.A. degrees in business administration from
Sungkyunkwan University, Seoul, South Korea, and the Ph.D. degree in
information and communication media engineering from Seoul National University
of Science and Technology, Seoul, South Korea.

He is currently a Professor with the Graduate School of AI Convergence
Technology and the Chair of the Department of AI Creation, College of AI
Convergence, Seoul Cyber University, Seoul, South Korea. He was previously an
Adjunct Professor with Sungkyunkwan University and Hanyang University and led
the Color Technology Laboratory for more than ten years. He has directed
industry--academic research projects with the Electronics and
Telecommunications Research Institute (ETRI), Hyundai-Kia Motor Group, and
Adobe Systems Korea. He teaches generative AI, prompt engineering, and
AI-based content creation, and his research interests include the reliability
of AI assessment, AI in education policy, and generative AI for creation
workflows.
\end{IEEEbiography}

\begin{IEEEbiography}[{\includegraphics[width=1in,height=1.25in,clip,keepaspectratio]{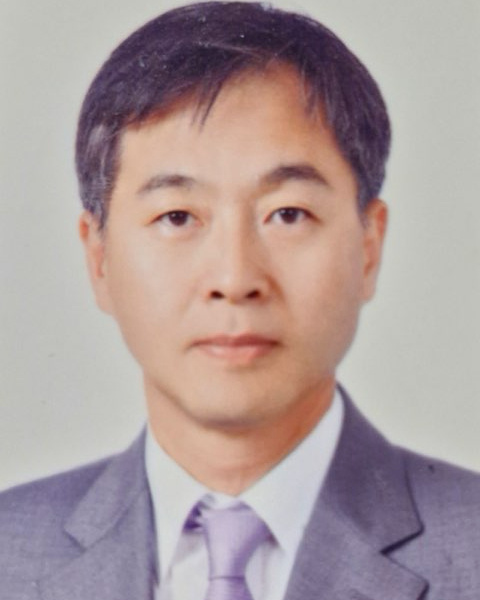}}]{Keuntae Cho}
received the Ph.D. degree in management of technology from Sungkyunkwan
University, Suwon, South Korea.

He is currently a Professor with the Department of Systems Management
Engineering and the Graduate School of Management of Technology, Sungkyunkwan
University. He served as the President of the Korea Society for Innovation
Management and Economics. His teaching and research interests include the
management of technology, technology commercialization, and the analytic
hierarchy process.
\EOD
\end{IEEEbiography}

\typeout{get arXiv to do 4 passes: Label(s) may have changed. Rerun}
\end{document}